# A Polynomial Time Algorithm for the Hamilton Circuit Problem

Xinwen Jiang .etc[**]

School of Computer, National University of Defense Technology,

ChangSha, P. R. China, 410073

**Abstract:** In this paper, we introduce a so-called 'Multistage graph **S**imple Path' (MSP) problem and show that the Hamilton Circuit (HC) problem can be polynomially reducible to the MSP problem. To solve the MSP problem, we propose a polynomial algorithm and prove its NP-completeness. Our result implies NP=P.

## 1. Introduction

The Hamilton Circuit problem is a well-known NP-complete problem [1]. This famous problem can be described as follows:

Given an undirected graph G=(V, E), does G have a Hamilton Circuit, i.e., a circuit visiting each vertex in V exactly once?

This problem has attracted a mount of attention since it was born. However, no polynomial time algorithm has been designed until now，nor has a proof that this problem can not be solved in polynomial time been confirmed. A piece of recent work made by HP Lab's Vinay Deolalikar has caused much discussion, debate and comment on the Internet. He claimed that he had proved P≠NP, but unfortunately, there are some flaws in his proof [11].

This paper presents the full version of our idea to solve this famous problem. We will introduce a so-called 'Multistage graph **S**imple Path' (MSP) problem and prove its NP-completeness. To solve the MSP problem, we will propose a polynomial algorithm and prove its correctness.

The following part of this paper includes 4 sections.

(1) MSP problem and definitions

(2) Z-H algorithm to solve MSP

(3) Proving MSP∈NPC

(4) Conclusions

## 2. MSP problem and definitions

We begin with defining a kind of multistage graph.

---

[**] Xinwen Jiang    http://trytoprovenpvsp.blog.sohu.com/170695074.html

**Definition 1** A *labeled multistage graph* G=<V, E, S, D, L> is a directed graph with the following properties:

(1) V is the vertex set of G. $V=V_0 \cup V_1 \cup V_2 \cup \ldots \cup V_L$, $V_i \cap V_j = \emptyset$, $0 \leq i, j \leq L$, $i \neq j$. If $u \in V_i$, $0 \leq i \leq L$, we say that u is a vertex of stage i, where L is the number of the stages of G. (Ø means empty in this paper)

(2) E is the edge set of G. Any edge in E is a directed one. We use <u, v, *l*> to represent an edge in E and say that <u, v, *l*> is an edge of stage *l*. If <u, v, *l*> ∈ E, then $u \in V_{l-1}$, $v \in V_l$, where, $1 \leq l \leq L$.

(3) $V_0$ and $V_L$ have only one vertex. The vertex in $V_0$ is named as S and the vertex in $V_L$ is named as D.

(4) Each vertex v ∈ V −{S} is labeled with E(v) that is a subset of E. We call E(v) as edge set of v.

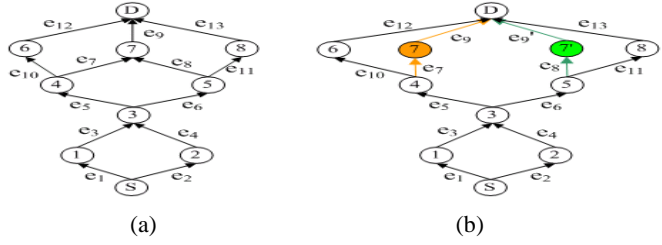

(a) (b)

Figure 1 Two Examples of labeled multistage graph

**Example 1** The two graphs shown in Fig.1 are both labeled multistage graphs. In Fig.1(a), E(1)={$e_1$}, E(2)={$e_2$}, E(3)={ $e_1$, $e_2$, $e_3$, $e_4$}, E(4) ={ $e_1$, $e_3$, $e_5$}, E(5) ={ $e_2$, $e_4$, $e_6$}, E(6)={ $e_1$, $e_3$, $e_5$, $e_{10}$}, E(7) ={$e_{12}$}, E(8) ={ $e_1$, $e_3$, $e_6$, $e_8$}, E(D)= { $e_1$, $e_3$, $e_5$, $e_{10}$, $e_{12}$}. In Fig.1(b), E(1)= Ø, E(2)= Ø, E(3)= Ø, E(4) ={ $e_1$, $e_3$, $e_5$}, E(5) ={ $e_2$, $e_4$, $e_6$}, E(6)={ $e_1$, $e_3$, $e_5$}, E(7) =E(7')= { $e_1$, $e_3$, $e_6$, $e_8$}, E(8) ={ $e_1$, $e_3$, $e_6$, $e_8$}, E(D) = Ø.

**Definition 2** Let G=<V, E, S, D, L> be a labeled multistage graph and S−$u_1$−…−$u_l$−…−$u_L$ ($1 \leq l \leq L$, $u_L$=D) be a path from S to D in G. S−$u_1$−…−$u_l$−…−$u_L$ is called a *simple path* in G if S−…−$u_l$ ∈ E($u_l$) for *l* ∈ {1, 2, …, L} . S−$u_1$−…−$u_l$−…−$u_L$ is called a pre-simple path in G if S−…−$u_l$ ∈ E($u_l$) for *l* ∈ {1, 2, …, L-2}.

For simplicity, we write S−…−v ∈ E(v) to denote that all the edges on S−…−v are contained in E(v). Now we propose a problem called 'Multistage graph Simple Path' (MSP) problem as follows.

**Given a labeled multistage graph G=<V, E, S, D, L>, does G have a simple path, i.e., a path S−…−v−…−D such that S−…−v ∈ E(v) for all v on S−…−v−…−D except S?**

Some labeled multistage graphs contain simple path and others do not. For example, there is a simple path S−1−3−4−6−D in Fig 1(a), while there is no simple path in Fig.1(b).

As we know, the result of the famous TSP problem may be different if we change the weights of some edges rather than the basic structure of G. It is the same case for MSP problem. If we change E(v), the existence of a simple path may be changed even if we do not make any change to the basic structure of G. For example, if we change the value of E(v) in Fig.1(a) into E(1)={$e_1$}, E(2)={$e_2$}, E(3)={ $e_1$, $e_2$, $e_3$, $e_4$}, E(4) ={ $e_1$, $e_3$, $e_5$}, E(5) ={ $e_2$, $e_4$, $e_6$}, E(6)={ $e_1$, $e_3$, $e_5$}, E(7) ={$e_{12}$}, E(8) ={ $e_1$, $e_3$, $e_6$, $e_8$}, E(D)= {$e_1$}, then the labeled multistage graph shown in Fig.1(a) has no simple path.

MSP problem is to determine the existence of a simple path in a labeled multistage graph. Obviously, this problem is a NP problem. We can solve this problem on any NDTM easily, but we will give an algorithm which is totally different from exhausting algorithm on DTM to solve MSP problem in this paper.

According to the definition of labeled multistage graph, an edge may appear in several E(v)'s. From the definition of simple path, if $S-u_1-\ldots-u_{l-1}-u_l-\ldots-D$ ($1 \leq l \leq L$, $u_0=S$, $u_L=D$) is a simple path, then $<u_{l-1}, u_l, l> \in E(u_l), E(u_{l+1}), \ldots, E(D)$.

**Definition 3** Let $G=<V, E, S, D, L>$ be a labeled multistage graph. If there exists a path $v-v_1-v_2-\ldots-v_k-\ldots-D$ in G such that $<u, v, l> \in E(v), E(v_1), E(v_2), \ldots, E(v_k), \ldots, E(D)$, we say that $v-v_1-v_2-\ldots-v_k-\ldots-D$ is a *reachable path* of $<u, v, l>$ from $E(v)$ to $E(D)$. All the reachable paths of $<u,v, l>$ form the reachable path set of $<u,v,l>$.

In this paper, we use R (u, v, *l*) to collect the edges on the reachable paths of $<u, v, l>$, and use R(E) to denote $\{R (u, v, l) | <u, v, l> \in E\}$. It is worthy noting that R(u,v,*l*) is a set of edges rather than paths.

Befor describing our algorithm to solve MSP problem, we firstly define four basic operators.

**Operator 1:** $[\text{ES}]_u^v$.

Suppose $ES \subseteq E$ and $u,v \in V$. We define $[\text{ES}]_u^v = \{e | e \in ES$, e is on a path $u-\ldots-v$, and all the edges on $u-\ldots-v$ are contained in ES$\}$.

$[\text{ES}]_u^v$ is defined to 'tidy' ES. Only those edges that are on some paths from u to v in ES are kept in $[\text{ES}]_u^v$. Let $|E|$ be the number of the edges in E. The computation of $[\text{ES}]_u^v$ can be finished in $O(|E|)$, since $|ES| \leq |E|$.

**Operator 2:** Init (R (u, v, *l*)).

Init (R (u, v, *l*)) is to compute the initial value of R( u, v, *l* ).

(1) $ES \leftarrow \{<a, b, k> | <a, b, k> \in E, l < k \leq L, <u, v, l> \in E(a) \cap E(b)\}$.   // Collecting edges

(2) $R(u, v, l) \leftarrow [\text{ES}]_v^D$.                            // Linking edges together

Let $|E|$ be the number of the edges in G. We can design an algorithm to compute Init(R(u, v, *l*)) in $O(|E|)$.

**Operator 3:** Comp(ES, v, R(E)).

Let ES be a subset of E, v be a vertex of V, $R(E)= \{ R (e) | e \in E\}$. Comp(ES, v, R(E)) equals the final result of ES_temp after the following iterations:

(1) ES_temp ← ES

(2) For all $e= <a, b, k> \in$ ES_temp:

　　　if $[R(a, b, k) \cap \text{ES\_temp}]_b^v$ contains no path from b to v, ES_temp ← ES_temp-{e}.

(3) ES_temp ← $[\text{ES\_temp}]_S^v$, where, S is the unique vertex of $V_0$.

(4) Repeat step 2 and step 3 until ES_temp will not change any more.

The second step of Comp(ES, v, R(E)) is to delete e if $[R(a, b, k) \cap ES\_temp]_b^v$ is empty. This is because that e in this case is not on a simple path which traverses vertex v. The third step is to bind all edges in ES_temp together. It will delete all those edges that are not on a path from S to v in ES_temp. For each vertex v which is on a simple path S－…－v－…－D, we have S－…－v $\in$ E(v), so, those edges that are not on a path from S to v in ES_temp can not be on such a simple path P that P traverses v and $[P]_S^v \in$ ES_temp. In one word, the result of Comp(ES, v, R(E)) is a subset of ES, such that for all <a, b, k> in Comp(ES, v, R(E)), <a, b, k> is on a path from S to v in Comp(ES, v, R(E)) and R(a, b, k) $\cap$ Comp(ES, v, R(E)) contains a path from b to v.

Each iteration reduces at least one edge in ES_temp and the number of the edges in ES_temp is no more than |E|. The execution of the iteration will stop eventually. It is worthy noting that the result of Comp(ES, v, R(E)) may be empty.

Let's analyze the time complexity of Comp(ES, v, R(E)). An algorithm of $O(|E|^2)$ can be designed to finish step 2. Therefore we can finish step 2 and step 3 in $O(|E|^2)$. The execution of Comp(ES, v, R(E)) will terminate before it reaches |E| iterations, since at least one edge is deleted during each iteration and the number of the edges in ES_temp is no more than |E|. Thus the complexity of Comp(ES, v, R(E)) is $O(|E|^3)$.

**Operator 4:** Change (R(u, v, *l*)).

Change(R(u, v, *l*)) is used to modify R(u, v, *l*). The key idea of this operator is to use R(E) to bind and limit R(u, v, *l*):

(1) For all <a, b, k> $\in$ R(u, v, *l*), *l* <k $\leq$ L

   **if** Comp([ {e | e=<c, d, kk> $\in$ E, kk < *l*, $[R(e) \cap Comp(E(b), b, R(E))]_d^b$ contains <u, v, *l*> and <a, b, k>} $]_S^u$, u, R(E)) $\neq \emptyset$

   **then** <a, b, k> is kept in R(u, v, *l*);

   **else** <a, b, k> is deleted from R(u, v, *l*).

(2) R(u, v, *l*) ← $[R(u, v, l)]_v^D$.

(3) Repeat step 1 and step 2 until R(u, v, *l*) will not change any more.

(When Operator 4 works, R(E) = { R (e) | e $\in$ E} should be a global variable. )

In step 1, if <a, b, k> is kept in R(u, v, *l*), there must exist a path S－…－u such that R(e) contains <u, v, *l*> and <a, b, k> for all e on S－…－u. Or more simply, <u, v, *l*> can pass across <a, b, k> only if there exists a path P=S－…－u such that P pass across <u, v, *l*> and <a, b, k>.

After step 2, R(u, v, *l*) only holds edges that are on the paths from v to D.

The time complexity of Change(R(u, v, *l*)) depends on operator 1, 2 and 3. We can get {e | e=<c, d, kk> $\in$ E, kk < *l*, $[R(e) \cap Comp(E(b), b, R(E))]_d^b$ contains <u, v, *l*> and <a, b, k>} in $|E|*O(|E|^3)$ and finish step 1 in

$|E|*|E|*O(|E|^3)$. The execution of Change(R(u, v, *l*)) will terminate before it reaches |E| iterations, since at least one edge is deleted during each iteration. Therefore, the complexity of Change(R(u, v, *l*)) is $|E|*|E|*|E|*O(|E|^3) = O(|E|^6)$.

After the execution of Change(R(u,v,*l*)), R(u, v, *l*) becomes a subset of its original value.

Before further discussion, we would like to point out again that all R(u, v, *l*) are subsets of E. Although we use R(E) to hold reachable paths, they only collect edges on these paths actually. This may bring in unexpected paths. However, these unexpected paths will not influence the determination of the existence of a simple path in our algorithm, and using edge sets to represent the reachable path sets can significantly reduce complexity.

## 3. Z-H algorithm to solve MSP

We now begin to prove that we can determine the existence of a simple path in a given multistage graph by a criterion that results from a series of modifications on R(u, v, *l*) and E(v).

### 3.1 Z-H algorithm, Complexity of Z-H algorithm and the proof of necessity

Let ES be an edge set. We use ES[i:j] to denote the set of all edges of ES from stage i to stage j, where $1 \leq i \leq j \leq L$. If i>j, ES[i:j] = Ø.

We propose the following so-called **Z-H algorithm** to solve the MSP problem. The input of the algorithm is G= <V, E, S, D, L>.

___________________________________________________________________________

1. For all e∈E, we use operator 2 to generate R(e) directly.

2. For *l*=1 to L-1

    2.1 For all <u, v, *l*> of stage *l*, call Change(R(u, v, *l*)) to modify R(u, v, *l*)

    2.2 For all v of stage *l*,  E(v) ← Comp(E(v), v, R(E))

    2.3 For all <a, b, k>∈E, k≤*l*, execute the following two steps:

    $$R(a, b, k)[k+1:l] \leftarrow \bigcup_{v \in V_l} [\ R(a, b, k) \cap Comp(E(v), v, R(E))\ ]_b^v \quad \text{// Limit R(e)}$$

    $$R(a, b, k) \leftarrow [\ R(a, b, k)\ ]_b^D \quad \text{//Tidy R(e)}$$

3. Repeat step 2 until no R(u, v, *l*) in R(E)= { R(e) | e∈E} will change any more.

4. If Comp(E(D), D, R(E))≠Ø, we claim the existence of a simple path in G. Otherwise, we claim that there is no simple path in G.

___________________________________________________________________________

According to the definition of notation ES[i:j], R(a, b, k)[k+1:*l*] in the above algorithm represents all edges of R(a, b, k) from stage k+1 to stage *l*. 'R(a, b, k)[k+1:*l*]← $\bigcup_{v \in V_l}$ [ R(a, b, k) ∩ Comp(E(v), v, R(E)) ]$_b^v$ ' means a part of R(a, b, k) is replaced by $\bigcup_{v \in V_l}$ [ R(a, b, k)∩Comp(E(v), v, R(E)) ]$_b^v$. Before the replacement, we have R(a, b, k)=[ R(a, b, k) ]$_b^D$. After the replacement, we need to 'tidy' R(a, b, k) to keep R(a, b, k)=[ R(a, b, k) ]$_b^D$ again.

The reason why we can do so strict a replacement here is that all edge sets and all reachable path sets on a simple are completely overlapped.

R(a, b, k) holds reachable paths of <a, b, k>. If we imagine an action to move <a, b, k> from E(b) to E(D) along the path in R(a, b, k), R(a, b, k) is actually the tracks of <a, b, k>. 'R(a, b, k)[k+1:*l*]←[ R(a, b, k)∩Comp(E(v), v, R(E)) ]$_b^v$' forces portion of these tracks of <a, b, k> to be kept in Comp(E(v), v, R(E)). '$\bigcup_{v \in V_l}$ [ R(a, b, k)∩Comp(E(v), v, R(E)) ]$_b^v$' collects all '[ R(a, b, k)∩Comp(E(v), v, R(E)) ]$_b^v$' together. Simply to say, if R(a, b, k) contain <u, v, *l*>, there must exist a path P=b—…—u—v such that P∈Comp(E(v), v, R(E)) and P∈R(a, b, k).

The main idea of Z-H algorithm is to use all the reachable path sets of stage 1, stage 2,…, stage *l*-1 to bind R(u, v, *l*) (step 2.1), use R(E) to modify the edge set E(v) (step 2.2), and use Comp(E(v), v, R(E)) to modify all the reachable path sets (step 2.3), one stage after another stage. After step 3, we use R(E) to compute Comp(E(D), D, R(E)). Comp(E(D), D, R(E))≠Ø means that Comp(E(D), D, R(E)) contains at least one path from S to D.

Our conclusion is amazingly simple: **G contains a simple path if and only if Comp(E(D), D, R(E))≠Ø**.

As you noticed, we use operator 1 to define operator 2, and then use operator 1 and 2 to define operator 3. After that, we use operator 1, 2, and 3 to define operator 4, and finally use operator 1, 2, 3, 4 to define Z-H algorithm.

**Theorem 1** Let |V| be the number of the vertices and |E| be the number of the edges in G. The time complexity of Z-H algorithm is a polynomial function of |V|*|E|.

**Proof:** Since |V| is the number of the vertices and |E| is the number of the edges in G, we can infer that the number of edges in each edge set and reachable path set is no more than |E|, the number of reachable path set (that is |R(E)|) is no more than |E|, and the number of E(v) (that is |{E(v)|v∈V-{S}}|) is no more than |V|.

The complexity for computing Comp(ES, v, R(E)) is $O(|E|^3)$ and the complexity for computing Change(R(u, v, *l*)) is $O(|E|^6)$, hence the complexity for step 2 of Z-H algorithm is $O(|E|^7)$.

Step 2 is the most complex statement in Z-H algorithm. Each iteration of step 2 will reduce at least one edge in R(u, v, *l*) and the number of edges in R(u, v, *l*) is no more than |E|. The number of R(e) is no more than |R(E)|. So, the complexity of step 2 and step 3 is |E| *|R(E)|* $O(|E|^7)$. This implies that the time complexity of Z-H algorithm is a polynomial function of |V|*|E|.    ∎

**Theorem 2** If there exists a simple path in G, then, we will get Comp(E(D), D, R(E))≠Ø after the execution of Z-H algorithm.

**Proof:** Let $v_0—v_1—v_2—…—v_L$ be a simple path in G, $v_0$=S, $v_L$ = D. According to the definition of simple path, $v_0—v_1—v_2—…—v_l$∈E($v_l$), 1≤*l*≤L, and, for <$v_{l-1}$, $v_l$, *l*> on $v_0—v_1—v_2—…—v_L$, 1≤*l*≤L, we have <$v_{l-1}$, $v_l$, *l*>∈E($v_l$), E($v_{l+1}$), E($v_{l+2}$), …, E(D). So, after the execution of the first step of Z-H algorithm, R($v_{l-1}$, $v_l$, *l*) will contain $v_l—v_{l+1}—…—v_L$, 1≤*l*≤L. After step 2, R($v_{l-1}$, $v_l$, *l*) will contain $v_l—v_{l+1}—…—v_L$, 1≤*l*≤L. Step 3 can not cut any path in R($v_{l-1}$, $v_l$, $l$). This will assure that Comp(E(D), D, R(E)) contains $v_0$—$v_1$—$v_2$—…—$v_L$. Hence, Comp(E(D), D, R(E))≠Ø. ∎

### 3.2 Getting ready to prove sufficiency

Can we claim the existence of a simple path in a given multistage graph if Comp (E(D), D, R(E)) ≠Ø?

In order to prove that the claim is correct, we need to introduce a metric to evaluate the "complexity" of a given multistage graph.

#### 3.2.1 The lexicographical order and its application in multistage graph

We define a lexicographical order "≤". For any two n-dimension vectors X = ($x_1$, $x_2$, …, $x_n$) and Y= ($y_1$, $y_2$, …, $y_n$) in $R^n$, if $(\exists k > 0)(\forall i < k)(x_i = y_i) \wedge (x_k < y_k)$, we say X<Y; if $x_i = y_i$, 1≤i≤n, we say X=Y.

Therefore, according to our definition, for any two vectors of n-dimension X = ($x_1$, $x_2$, …, $x_n$) and Y= ($y_1$, $y_2$, …, $y_n$) in $R^n$, we have X≤Y or Y≤X.

For each multistage graph G, we define the following vector for G:

Vec (G) = ($x_1$, $x_2$,…, $x_{L-1}$, $x_L$), where,

$$x_l = \sum_{v \in V_l of G}(d(v)-1), 1 \leq l \leq L-1,$$

$$x_L = 0,$$

d(v) is the in-degree of v, $V_l$ is the set of all the vertices of stage $l$.

**Definition 4** Let G=<V, E, S, D, L> and $G_1$=<$V_1$, $E_1$, S, D, L> be two multistage graphs. If Vec(G)<Vec($G_1$), we say that G<$G_1$.

#### 3.2.2 Two functions

We need to introduce a renaming function $I_y^x$ and a splitting function $I_v^{v_1,v_2}$. These two functions are defined to deal with triples. A triple in a multistage graph is an edge.

**Function $I_y^x$**: Let EL be a set, ET={<a, b, k> | a, b∈EL, k is an integer}, ES⊆ET, e∈ET, and x, y∈EL. $I_y^x$ is defined as follows.

$$I_y^x(\{e\}) = \begin{cases} \{<b,y,k>\}, \text{if } e=<b,x,k> \\ \{<y,b,k>\}, \text{if } e=<x,b,k> \\ \{e\}, \text{otherwise} \end{cases}$$

$$I_y^x(ES) = \bigcup_{e \in ES} I_y^x(\{e\})$$

**Function $I_v^{v_1,v_2}$**: Let EL be a set, ET={<a, b, k> | a, b∈EL, k is an integer}; v, $v_1$, $v_2$∈EL, $l$ be an integer; ES, $ES_1$, $ES_2$ be subsets of ET, $ES_1$≠Ø, $ES_2$≠Ø, $ES_1 \cap ES_2$=Ø, $ES_1 \cup ES_2$={e | e∈ES, e=<c, v, $l$>, c∈EL}. $I_v^{v_1,v_2}$ is

defined as follows.

$$I_v^{v_1,v_2}(ES, ES_1, ES_2) = (ES - \{e \mid e \in ES, e = <a,v, l> \text{ or } e = <v, a, l+1>, a \in EL\}) \cup$$

$$\{e \mid e = <a, v_1, l>, <a, v, l> \in ES_1, a \in EL\} \cup \{e \mid e = <a, v_2, l>, <a, v, l> \in ES_2, a \in EL\} \cup$$

$$\{e \mid e = <v_1, a, l+1> \text{ or } e = <v_2, a, l+1>, <v, a, l+1> \in ES, a \in EL\}.$$

In the splitting function $I_v^{v_1,v_2}$, we delete all triples with form $<a, v, l>$ or $<v, a, l+1>$ from ES, use $e=<a, v_1, l>$ to substitute $<a, v, l>$ if $<a, v, l> \in ES_1$, use $e=<a, v_2, l>$ to substitute $<a, v, l>$ if $<a, v, l> \in ES_2$, and use $e=<v_1, a, l+1>$ and $e=<v_2, a, l+1>$ to substitute $<v, a, l+1>$ if $<v, a, l+1> \in ES$.

### 3.2.3 Defining a proving algorithm

Why do we define a new algorithm before finishing the proof of Z-H algorithm?

If we want to prove that f(x) has some properties, we may turn to prove that $\frac{f(x)}{\sin^2(x) + \cos^2(x)}$ has some properties. This is a way that we often used to do mathematical proof. The reason why we choose this way to do mathematical proof is that we find it difficult to do proof directly. Here we meet the same problem: we find it difficult to prove the sufficiency of Z-H algorithm and therefore we discuss a more general case.

Based on Z-H algorithm, we design a new algorithm named as **Proving Algorithm**. To describe the algorithm, we need two symbols.

$/e_1/e_2/.../e_{|E_2|}$: We use $/e_1/e_2/.../e_{|E_2|}$ to indicate an edge set that holds some edges at stage 2 in G.

$R_{ini}(a, b, L-2)$: $R_{ini}(a, b, L-2)$ is a subset of E and $R_{ini}(a, b, L-2)$ is independent of $R(a, b, L-2)$ in the Proving Algorithm, where a, b $\in$ V, $<a, b, L-2> \in$ E.

**Here is the Proving Algorithm**, the inputs of the Proving Algorithm include $/e_1/e_2/.../e_{|E_2|}$, $\{R_{ini}(a, b, L-2) \mid <a, b, L-2> \in E\}$, G=<V, E, S, D, L> and an edge set ESS (ESS is a subset of E).

______________________________________________________________________

1. For all e∈E, we use operator 2 to generate R(e) directly.
2. For *l*=1 to L-2

    2.1 For all <u, v, *l*> of stage *l*, call Change(R(u, v, *l*)) to modify R(u, v, *l*), where k⩽L-2 for all <a, b, k> in step 1 of Change(R(u, v, *l*))

    2.2 For all v of stage *l*,  E(v) ← Comp(E(v), v, R(E))

    2.3 For all <a, b, k>∈E，k⩽*l*, execute the following two steps：

$$R(a, b, k)[k+1: l] \leftarrow \bigcup_{v \in V_l} [\ R(a, b, k) \cap Comp(E(v), v, R(E))\ ]_b^v \qquad // \text{Limit } R(e)$$

$$R(a, b, k) \leftarrow [\ R(a, b, k)\ ]_b^D \qquad // \text{Tidy } R(e)$$

3. Repeat step 2 until no R(u, v, *l*) in R(E) will change any more.

4. For all w of stage L-1, we check condition a) and b) as follows.

    a) E(w) ={<u, w, L-1>}∪Comp(E(u), u, R(E)), where u is a vertex at stage L-2,

    b) E(w)[L-1: L-1]=Ø,

    If there exists such a vertex w that both a) and b) are false, set all R(e) empty.

5. ESS1←ESS∩Comp(E(D), D, R(E)).

6. For all v∈V, recover E(v) to hold the value before step 1.

7. For all v of stage L-2, E(v) ←E(v)∪E[3:L]; For all w of stage L-1, E(w) ←E(w)∪E.

8. Execute step 1,2,3 again.

9. For all v∈V, recover E(v) to hold the value before step 1.

10. If we have:

    (1) Comp(ESS1, D, R(E))≠Ø (we call this condition as $H_1$),

    (2) $/e_1/e_2/.../e_{|E_2|}$ ∩ Comp(ESS1, D, R(E)) contains <aa, bb, 2>, such that $[\ R(aa, bb, 2) \cap Comp(ESS1, D, R(E))\ ]_{bb}^D$ contains bb– $b_3$ – $b_4$ – …– $b_{L-3}$– $b_{L-2}$– $b_{L-1}$–D, in which $b_{L-2}$– $b_{L-1}$– D is contained in $R_{ini}(b_{L-3}, b_{L-2}, L-2)$ (we call this condition as $H_2$),

    Then:

    there exists a simple path P=S – $a_1$ – $a_2$ – …– $a_{L-3}$– $a_{L-2}$– $a_{L-1}$– D in G, such that

    a) ESS contains P,

    b) < $a_1$, $a_2$, 2> is in $/e_1/e_2/.../e_{|E_2|}$,

    c) $R_{ini}(a_{L-3}, a_{L-2}, L-2)$ contains $a_{L-2}$– $a_{L-1}$– D

---

Let's have a look at the difference between Z-H algorithm and the Proving Algorithm firstly.

Briefly speaking, the Proving Algorithm repeats Z-H algorithm twice. The first time is from step 1 to step 5 and the second time is from step 6 to step 10. We get ESS∩ Comp(E(D), D, R(E)) in step 5, and after that, repeat step 1, 2, 3 to compute Comp(ESS1, D, R(E)). We have ESS1 $\subseteq$ ESS in the algorithm.

The Proving Algorithm has step 4, 5, 6, 7, 8, 9, while Z-H algorithm does not contain these steps. According to step 4, if we finally have Comp(ESS1, D, R(E)) ≠ Ø in step 10, for all w of stage L-1, we must have E(w) ={<u, w, L-1>}$\cup$Comp(E(u), u, R(E)), or, E(w)[L-1：L-1]= Ø. After step 4, we get ESS1. Then, we expand all E(v) of stage L-2 and stage L-1, and execute step 1,2,3 again. Such expansion is very important to the proof of lemma 4. Step 9 recovers all the values of E(v) that have been changed after step 6, since we assert the existence of a simple path in step 10, and the simple path depends on the value of all E(v). After step 9, all E(v) keep their initial value. Recall that the simple path we assert in the Proving Algorithm depends on the initial values of all E(v).

It is important to point out that the simple path claimed in step 10 of the Proving Algorithm depends on the input graph, $R_{ini}(a,b,L-2)$, $/e_1/e_2/.../e_{|E_2|}$ and ESS, while the simple path claimed in step 4 of Z-H algorithm depends only on the input graph.

We are inspired by the following facts (we will use these facts in the proof of theorem 3)**:**

(1) If $|V_{L-2}|= |V_{L-1}|= |V_L|=1$, the expansion of step 7 may become meaningless.

(2) If ESS=E(D), step 5 may become meaningless.

(3) If the given graph has only one edge at stage 2, $|V_{L-2}|= |V_{L-1}|= |V_L|=1$, $R_{ini}(a,b,L-2) \supseteq [E[L-1:L]]_b^D$, and ESS=E(D), step 10 in the Proving Algorithm may become step 4 in Z-H algorithm.

### 3.3　αβ lemma and its proof

We begin to prove that the claim of the **Proving Algorithm** is correct.

**Lemma 1.** Let G = <V, E, S, D, L> be the input of the Proving Algorithm, and, no multi-degree vertex can be found from stage 1 to stage L-1 in G (shown in Fig.4). After applying the Proving Algorithm on G, if we have

(1) Comp(ESS1, D, R(E))≠Ø,

(2) $/e_1/e_2/.../e_{|E_2|}$ $\cap$ Comp(ESS1, D, R(E)) contains <aa, bb, 2>, such that [ R(aa, bb, 2)$\cap$Comp(ESS1, D, R(E)) $]_{bb}^D$ contains bb– $b_3$ – $b_4$ – …– $b_{L-3}$– $b_{L-2}$– $b_{L-1}$–D, in which $b_{L-2}$– $b_{L-1}$– D is contained in $R_{ini}(b_{L-3},b_{L-2},L-2)$,

then, G has a simple path P=S – $a_1$ – $a_2$ – …– $a_{L-3}$– $a_{L-2}$– $a_{L-1}$– D, such that ESS contains P, < $a_1$, $a_2$, 2> is in $/e_1/e_2/.../e_{|E_2|}$, and $R_{ini}(a_{L-3},a_{L-2},L-2)$ contains $a_{L-2}$– $a_{L-1}$– D.

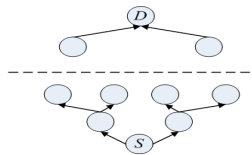

Figure 4 Typical graph of lemma 1

**Proof:** According to the condition (2), $/e_1/e_2/.../e_{|E_2|} \cap \text{Comp}(\text{ESS1}, D, R(E))$ contains <aa, bb, 2>, such that $[R(aa, bb, 2) \cap \text{Comp}(\text{ESS1}, D, R(E))]_{bb}^{D}$ contains $bb- b_3 - b_4 - ...- b_{L-3}- b_{L-2}- b_{L-1}-D$, in which $b_{L-2}- b_{L-1}- D$ is contained in $R_{ini}(b_{L-3}, b_{L-2}, L-2)$, hence we have:

(1) ESS contains $S-aa-bb- b_3 - b_4 - ...- b_{L-3}- b_{L-2}- b_{L-1}-D$.

(2) <aa, bb, 2> is in $/e_1/e_2/.../e_{|E_2|}$.

(3) $R_{ini}(b_{L-3}, b_{L-2}, L-2)$ contains $b_{L-2}- b_{L-1}- D$.

(4) ESS∩Comp(E(D), D, R(E)) contains $b_{L-3}- b_{L-2}- b_{L-1}-D$ in step 5. This implies that the initial $E(b_{L-1})$ equals $\{< b_{L-2}, b_{L-1}, L-1>\} \cup \text{Comp}(E(b_{L-2}), b_{L-2}, R(E))$ and $\text{Comp}(E(b_{L-2}), b_{L-2}, R(E))$ is not empty. Therefore, the initial $E(b_{L-2})$ and the initial $E(b_{L-1})$ must contain $S-aa-bb- b_3 - b_4 - ...- b_{L-3}- b_{L-2}$. Thus we know that $S-aa-bb- b_3 - b_4 - ...- b_{L-3}- b_{L-2}- b_{L-1}-D$ is a simple path in G. ∎

Now we begin to prove the Proving Algorithm is correct for all graphs.

**We can directly verify that the Proving Algorithm can make correct assertion for all multistage graphs with four stages. So, in the following discussion, we assume that the Proving Algorithm can make correct assertion for all multistage graphs with L-1 stages. For all graphs with L stages, if the Proving Algorithm is incorrect for some multistage graphs of this kind, we can find out the 'smallest' one with respect to the linear order "≤" which we defined above. Therefore, without losing generality, we can further assume that G is the 'smallest' graph of this kind that makes the Proving Algorithm fail in determining the existence of the said simple path in step 10.**

According to the definition of labeled multistage graph, an edge set E(x) is a subset of E. However, according to the definition of simple path, only those edges in $[E(x)]_S^x$ could be on the simple paths that traverse x. And actually, all edges in $E(x)-[E(x)]_S^x$ will certainly be deleted in step 2.2 of the Proving Algorithm. Therefore, in the following discussions of Lemma 2,3,4,5, we assume that each initial E(x) in G equals $[E(x)]_S^x$ when G is the input of the Proving Algorithm. However, when we construct a new graph and assign a value to its E(x), we do not require that E(x) equals $[E(x)]_S^x$ for the sake of convenience.

**Lemma 2.** Let G = <V, E, S, D, L> be the input of the Proving Algorithm, vertex v of stage L-1 is a multi in-degree vertex, as shown in Fig.5(a). After applying the Proving Algorithm on G, if we have

(1) Comp(ESS1, D, R(E))≠Ø,

(2) $/e_1/e_2/.../e_{|E_2|} \cap \text{Comp}(\text{ESS1}, D, R(E))$ contains <aa, bb, 2>, such that $[R(aa, bb, 2) \cap \text{Comp}(\text{ESS1}, D, R(E))]_{bb}^{D}$ contains $bb- b_3 - b_4 - ...- b_{L-3}- b_{L-2}- b_{L-1}-D$, in which $b_{L-2}- b_{L-1}- D$ is contained in $R_{ini}(b_{L-3}, b_{L-2}, L-2)$,

then, there must exist a simple path P=S – $a_1$ – $a_2$ – …– $a_{L-3}$– $a_{L-2}$– $a_{L-1}$– D in G, such that ESS contains P, < $a_1$, $a_2$, 2> is in $/e_1/e_2/.../e_{|E_2|}$, and $R_{ini}(a_{L-3}, a_{L-2}, L-2)$ contains $a_{L-2}$– $a_{L-1}$– D.

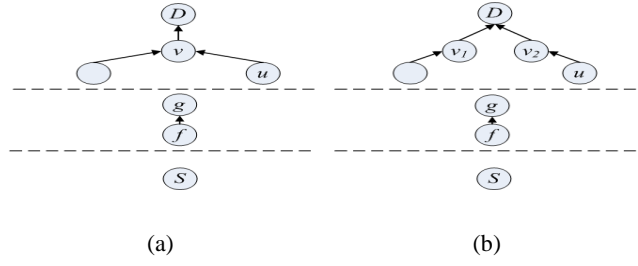

(a)        (b)

Figure 5 Typical graph of lemma 2

**Proof:** We summarize the main idea of the proof at first. To prove lemma 2, we need to construct a graph $G_1$, such that:

(1) $G_1 < G$;

(2) If we have $H_1 \wedge H_2$ ($H_1 \wedge H_2$ means $H_1$ and $H_2$) when G is the input of the Proving Algorithm, we must have $H_1 \wedge H_2$ when $G_1$ is the input of the Proving Algorithm.

(3) If P is a simple path in $G_1$, P must be a simple path in G.

Proof begins. First, we construct a graph $G_1$=<$V_1$, $E_1$, S, D, L> as follows:

We arbitrarily separate all edges ending at v into two non-empty parts, $group_1$ and $group_2$, and split v into $v_1$ and $v_2$ in a bottom-up way. Thus we get a new multistage graph $G_1$ =<$V_1$, $E_1$, S, D, L> shown in Fig. 5(b).

(1) Vertices of $G_1$: $V_1$=(V-{v})∪{$v_1$,$v_2$}.

(2) Edges of $G_1$:

For all <u, v, L-1> in G, we generate <u, $v_1$, L-1> in $G_1$ if <u, v, L-1> ∈ $group_1$, or <u, $v_2$, L-1> in $G_1$ if <u, v, L-1>∈ $group_2$.

If <v, D, L> is an edge in G, we generate <$v_1$, D, L> and <$v_2$, D, L> in $G_1$.

All edges in G that do not start from v or end at v become edges in $G_1$.

(3) Edge set of $G_1$:

If E(v)={<u, v, L-1>}∪Comp(E(u), u, R(E)) and <u, v, L-1>∈ $group_1$, we set E($v_1$)= {<u, $v_1$, L-1>} ∪Comp(E(u), u, R(E)). If E(v)={<u, v, L-1>}∪Comp(E(u), u, R(E)) and <u, v, L-1>∈ $group_2$, we set E($v_1$)= Comp(E(u), u, R(E)) (Please note**:** <u, $v_1$, L-1> is not in E($v_1$) here. Hence it will satisfy the conditions of step 4 in the Proving Algorithm). We treat E($v_2$) in the same way as we treat E($v_1$).

If E(v)[L-1:L-1]=Ø, set E($v_1$)=E($v_2$)=E(v)[1:L-2].

For all x∈((V$_1$-{ D, $v_1$, $v_2$}) of $G_1$), set (E(x) of $G_1$)=(E(x) of G). (Please note: we have assumed that

E(x) equals $[E(x)]_S^x$ in G and we do not require that E(x) equals $[E(x)]_S^x$ in $G_1$)

Set (E(D) of $G_1$)= $I_v^{v_1,v_2}$ ((E(D) of G), $group_1$, $group_2$)

(4) ESS of $G_1$: Set (ESS of $G_1$)= $I_v^{v_1,v_2}$ ( (ESS of G), $group_1$, $group_2$).

(5) $/e_1/e_2/.../e_{|E_2|}$ of $G_1$: Set ($/e_1/e_2/.../e_{|E_2|}$ of $G_1$)= ($/e_1/e_2/.../e_{|E_2|}$ of G).

(6) $R_{ini}(a,b,L-2)$ of $G_1$: Set ($R_{ini}(a,b,L-2)$ of $G_1$) = $I_v^{v_1,v_2}$ ($R_{ini}(a,b,L-2)$ of G, $group_1$, $group_2$).

We explain some notations at first.

($/e_1/e_2/.../e_{|E_2|}$ of $G_1$) means $/e_1/e_2/.../e_{|E_2|}$ in $G_1$, and ($/e_1/e_2/.../e_{|E_2|}$ of G) means $/e_1/e_2/.../e_{|E_2|}$ in G. In this paper, we have to compare a quantity in G with its peer in $G_1$ frequently. For simplicity, we use the phrase like 'something of G' to express a quantity in G or in the Proving Algorithm when G is the input. $R_{ini}(a,b,L-2)$ of G and $R_{ini}(a,b,L-2)$ of $G_1$ are examples of this form. Sometimes, for simplicity and clarity, we use '(something of G)' to express a quantity in G furthermore. (E(x) of $G_1$), (E(x) of G), and ($/e_1/e_2/.../e_{|E_2|}$ of $G_1$) are examples of this form.

Go back to the proof. We prove conclusion (1), (2) and (3).

(1) $G_1 < G$.

v is the multi in-degree vertex which appears at stage $l$, $l$=L-1, Vec ($G_1$)[1:(L-2)]= Vec (G) [1: (L-2)].

$$
\begin{aligned}
\text{Vec } (G_1)(l) &= \sum_{u \in V_l \text{ of } G_1}(d(u)-1) \\
&= \sum_{u \in V_l-\{v_1,v_2\} \text{ of } G_1}(d(u)-1) + (d(v_1)-1) + (d(v_2)-1) \\
&= \sum_{u \in V_l-\{v_1,v_2\} \text{ of } G_1}(d(u)-1) + (d(v_1)+d(v_2)) - 2 \\
&= \sum_{u \in V_l-\{v_1,v_2\} \text{ of } G_1}(d(u)-1) + (d(v)-1) - 1 \\
&= \sum_{u \in V_l \text{ of } G}(d(u)-1) - 1 \\
&= \text{Vec } (G)(l)-1
\end{aligned}
$$

< Vec (G) (*l*)

Hence Vec(G$_1$)<Vec(G), and therefore, G$_1$<G.

(2) When G$_1$ is the input of the Proving Algorithm, we will have: (a) (Comp(ESS1, D, R(E)) of G$_1$)≠Ø, and (b) $/e_1/e_2/.../e_{|E_2|}$ ∩ (Comp(ESS1, D, R(E)) of G$_1$) contains <aa, bb, 2>, such that [ (R(aa, bb, 2) of G$_1$)∩ (Comp(ESS1, D, R(E)) of G$_1$) ]$_{bb}^{D}$ contains bb– b$_3$ – b$_4$ – …– b$_{L-3}$– b$_{L-2}$– b$_{L-1}$–D, in which b$_{L-2}$– b$_{L-1}$– D is contained in ( $R_{ini}(b_{L-3}, b_{L-2}, L-2)$ of G$_1$).

The reason is as follows.

Before step 5 of the Proving Algorithm:

For all vertices x at stage k, k<L-1, (Comp(E(x), x, R(E)) of G) ⊆(Comp(E(x), x, R(E)) of G$_1$).

For all <a, b, k> (k<L-1) and for all <a, b, L-1> (b≠v), if (R(a, b, k) of G) contains a path b– b$_{k+1}$ – b$_{k+2}$ – …– b$_{L-3}$– b$_{L-2}$– b$_{L-1}$–D, (R(a, b, k) of G$_1$) contains b– c$_{k+1}$ – c$_{k+2}$ – …– c$_{L-3}$– c$_{L-2}$– c$_{L-1}$–D such that $I_v^{v_2} I_v^{v_1}$ (b– c$_{k+1}$ – c$_{k+2}$ – …– c$_{L-3}$– c$_{L-2}$– c$_{L-1}$–D)= b– b$_{k+1}$ – b$_{k+2}$ – …– b$_{L-3}$– b$_{L-2}$– b$_{L-1}$–D.

For all <u, v, L-1> in G, if (R(u, v, L-1) of G) contains v–D and <u, v, L-1>∈group$_1$, then (R(u, v$_1$, L-1) of G$_1$) contains v$_1$–D; if (R(u, v, L-1) of G) contains v–D and <u, v, L-1>∈group$_2$, then (R(u, v$_2$, L-1) of G$_1$) contains v$_2$–D.

Therefore, ((ESS of G)∩Comp(E(D), D, R(E))) of G⊆ $I_v^{v_2} I_v^{v_1}$ ((ESS of G$_1$)∩Comp(E(D), D, R(E)) of G$_1$)

After step 5 of the Proving Algorithm:

For all vertices x at stage k, k<L-1, (Comp(E(x), x, R(E)) of G) ⊆(Comp(E(x), x, R(E)) of G$_1$)

For all <a, b, k> (k<L-1) and for all <a, b, L-1> (b≠v), if (R(a, b, k) of G) contains a path b– b$_{k+1}$ – b$_{k+2}$ – …– b$_{L-3}$– b$_{L-2}$– b$_{L-1}$–D, (R(a, b, k) of G$_1$) contains b– c$_{k+1}$ – c$_{k+2}$ – …– c$_{L-3}$– c$_{L-2}$– c$_{L-1}$–D such that $I_v^{v_2} I_v^{v_1}$ (b– c$_{k+1}$ – c$_{k+2}$ – …– c$_{L-3}$– c$_{L-2}$– c$_{L-1}$–D)= b– b$_{k+1}$ – b$_{k+2}$ – …– b$_{L-3}$– b$_{L-2}$– b$_{L-1}$–D.

For all <u, v, L-1> in G, if (R(u, v, L-1) of G) contains v–D and <u, v, L-1>∈group$_1$, then (R(u, v$_1$, L-1) of G$_1$) contains v$_1$–D; if (R(u, v, L-1) of G) contains v–D and <u, v, L-1>∈group$_2$, then (R(u, v$_2$, L-1) of G$_1$) contains v$_2$–D.

Therefore, (Comp(ESS1, D, R(E)) of G)⊆ $I_v^{v_2} I_v^{v_1}$ (Comp(ESS1, D, R(E)) of G$_1$).

Noting the facts that $(/e_1/e_2/.../e_{|E_2|}$ of G$_1$)= $(/e_1/e_2/.../e_{|E_2|}$ of G) and ( $R_{ini}(a,b,L-2)$ of G$_1$) = $I_v^{v_1,v_2}$ ( $R_{ini}(a,b,L-2)$ of G, group$_1$, group$_2$), we have $H_1 \wedge H_2$ when G$_1$ is the input of the Proving Algorithm.

(3) There exists a simple path P=S – a$_1$ – a$_2$ – …– a$_{L-3}$– a$_{L-2}$– a$_{L-1}$– D in G, such that ESS contains P, < a$_1$, a$_2$, 2>

is in $/e_1/e_2/.../e_{|E_2|}$, and $R_{ini}(a_{L-3}, a_{L-2}, L-2)$ contains $a_{L-2} - a_{L-1} - D$.

Since we have assumed that G is the "smallest" graph that fail the Proving Algorithm, and we have proved that $G_1 < G$, and we have $H_1 \wedge H_2$ when applying the Proving Algorithm on $G_1$, hence there must exist a simple path $P = S - a_1 - a_2 - ... - a_{L-3} - a_{L-2} - a_{L-1} - D$ in $G_1$, such that (ESS of $G_1$) contains P, $<a_1, a_2, 2>$ is in ($/e_1/e_2/.../e_{|E_2|}$ of $G_1$), and ($R_{ini}(a_{L-3}, a_{L-2}, L-2)$ of $G_1$) contains $a_{L-2} - a_{L-1} - D$.

If there exists a simple path $P = S - a_1 - a_2 - ... - a_{L-3} - a_{L-2} - a_{L-1} - D$ in $G_1$, such that (ESS of $G_1$) contains P, $<a_1, a_2, 2>$ is in ($/e_1/e_2/.../e_{|E_2|}$ of $G_1$), and ($R_{ini}(a_{L-3}, a_{L-2}, L-2)$ of $G_1$) contains $a_{L-2} - a_{L-1} - D$, then, $P' = I_v^{v_2} I_v^{v_1} (S - a_1 - a_2 - ... - a_{L-3} - a_{L-2} - a_{L-1} - D)$ must be a simple path in G, such that (ESS of G) contains P', $<a_1, a_2, 2>$ is in ($/e_1/e_2/.../e_{|E_2|}$ of G), and ($R_{ini}(a_{L-3}, a_{L-2}, L-2)$ of G) contains $I_v^{v_2} I_v^{v_1} (a_{L-2} - a_{L-1} - D)$. ∎

**Lemma 3.** Let $G = <V, E, S, D, L>$ be the input of the Proving Algorithm, vertex v of stage L-2 is a multi in-degree vertex, and no multi in-degree vertex can be found at stage L-1, as shown in Fig.6(a). After applying the Proving Algorithm on G, if we have

(1) Comp(ESS1, D, R(E))≠Ø,

(2) $/e_1/e_2/.../e_{|E_2|}$ ∩ Comp(ESS1, D, R(E)) contains $<aa, bb, 2>$, such that $[R(aa, bb, 2) \cap \text{Comp(ESS1, D, R(E))}]_{bb}^D$ contains $bb - b_3 - b_4 - ... - b_{L-3} - b_{L-2} - b_{L-1} - D$, in which $b_{L-2} - b_{L-1} - D$ is contained in $R_{ini}(b_{L-3}, b_{L-2}, L-2)$,

then, there must exist a simple path $P = S - a_1 - a_2 - ... - a_{L-3} - a_{L-2} - a_{L-1} - D$ such that ESS contains P, $<a_1, a_2, 2>$ is in $/e_1/e_2/.../e_{|E_2|}$, and $R_{ini}(a_{L-3}, a_{L-2}, L-2)$ contains $a_{L-2} - a_{L-1} - D$.

**Proof:** We summarize the main idea of the proof at first. To prove lemma 3, we need to construct a graph $G_1$ such that:

(1) $G_1 < G$;

(2) If we have $H_1 \wedge H_2$ when G is the input of the Proving Algorithm, we must have $H_1 \wedge H_2$ when $G_1$ is the input of the Proving Algorithm.

(3) If P is a simple path in $G_1$, P must be a simple path in G.

Proof begins. We arbitrarily separate all edges ending at v into two non-empty parts, namely, $group_1$ and $group_2$, and use a bottom-up way to split v into $v_1$, $v_2$, just as we split G in lemma 2 (please note: many multi in-degree vertices at stage L-1 will appear after the splitting). After that, we split all vertices at stage L-1 one by one, so that we can get a graph without multi in-degree vertex at stage L-1. Thus we get a new multistage graph $G_1$ shown by Fig. 6(b).

Set $V_1 = (V - \{x \mid x \in V-\{D\},$ x is a vertex on a path v– w– D in G$\}) \cup \{v_1, v_2\} \cup \bigcup_{For all v-w-D\ in\ G} \{w_1, w_2 \mid w_1, w_2$ are vertices at stage L-1, G contains v– w– D$\}$.

Set $E_1 = \bigcup_{\{w \mid d(w)\ of\ stage\ L-1 > 1,\ after\ splitting\ v\}} (I_w^{w_1,w_2}\ (I_v^{v_1,v_2}\ $(E of G, group$_1$, group$_2$), $\{<v_1, w, L-1>\}, \{<v_2, w, L-1>\}))$

Set $(/e_1/e_2/.../e_{|E_2|}$ of $G_1) = (/e_1/e_2/.../e_{|E_2|}$ of G); Set ($R_{ini}(a,b,L-2)$ of $G_1) = \bigcup_{\{w \mid d(w)\ of\ stage\ L-1 > 1,\ after\ splitting\ v\}}$

$(I_w^{w_1,w_2}\ (I_v^{v_1,v_2}\ (R_{ini}(a,b,L-2)$ of G, group$_1$, group$_2$), $\{<v_1, w, L-1>\}, \{<v_2, w, L-1>\}))$; Set (ESS of $G_1)=$

$\bigcup_{\{w \mid d(w)\ of\ stage\ L-1 > 1,\ after\ splitting\ v\}} (I_w^{w_1,w_2}\ (I_v^{v_1,v_2}\ $(ESS of G, group$_1$, group$_2$), $\{<v_1, w, L-1>\}, \{<v_2, w, L-1>\}))$.

Set (E(D) of $G_1) = \bigcup_{\{w \mid d(w)\ of\ stage\ L-1 > 1,\ after\ splitting\ v\}} (I_w^{w_1,w_2}\ (I_v^{v_1,v_2}\ $(E(D) of G, group$_1$, group$_2$), $\{<v_1, w, L-1>\}$, $\{<v_2, w, L-1>\}))$

All vertices, except those that are not D and originally on v– w– D before splitting, have the same edge sets as they have in G, namely, (E(x) of $G_1$)= (E(x) of G). We assign values to those vertices that are not D and originally on v– w– D before splitting later.

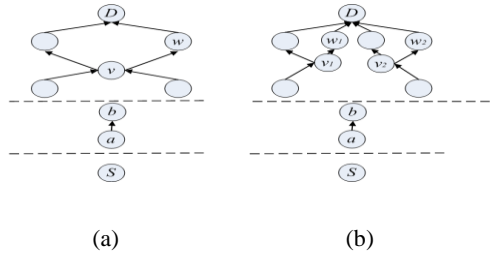

(a)      (b)

Figure 6 Typical graph of lemma 3

We can prove the following conclusion (1), (2), (3) and (4).

(1) $G_1 < G$.

The reason is similar to the proof in lemma 2.

(2) Before step 5, if (R(a, b, k) of G) contains <u, v, L-2> and v is the multi in-degree vertex that is going to be splitted, G must have a pre-simple path which traverses <a, b, k> and <u, v, L-2>.

(We suggest you to remember conclusion (2) and jump over its proof before you understand the main idea of the proof of Lemma 3.)

Since the Proving Algorithm does not compute Comp(E(v), v, R(E)) (v∈$V_{L-1}$), it is easy to assign values to E($v_1$) and E($v_2$) in Lemma 2. But we have to answer the following questions before splitting here. Dose (Comp(E(v), v, R(E)) of G) equals $(I_v^{v_2} I_v^{v_1}$ (Comp(E($v_1$), $v_1$, R(E)) of $G_1) \cup I_v^{v_2} I_v^{v_1}$ (Comp(E($v_2$), $v_2$, R(E)) of $G_1$)) or not? Does this splitting operation affect each R(e) or not? And, can we split v with holding $H_1 \wedge H_2$? The answer for each questions is 'yes' if conclusion (2) can be proved.

Proof for conclusion (2) begins. We are going to construct $G_2$ that is smaller than $G$, in which the simple path claimed by step 10 of the Proving Algorithm implies the existence of the pre-simple path that traverses <a, b, k> and <u, v, L-2> in G. Without losing generality, we assume <u, v, L-2>∈$group_1$.

Based on $G_1$, we construct $G_2$ by changing some values in $G_1$ as follows.

Set $V_2=V_1$. Set $E_2=E_1$.

Set $E(v_1)=E(v_2)=I_v^{v_1,v_2}$ ((E of G)[1:L-2], $group_1$, $group_2$) (please note: we assume that each initial E(x) in G equals $[E(x)]_S^x$ when G is the input of the Proving Algorithm. However, when we construct a new graph and assign a value to E(x), we do not require that E(x) equals $[E(x)]_S^x$ for the sake of convenience).

For all vertices w at stage L-1, if w is on $v_i$–w–D (i=1,2) in $G_2$, set E(w) equals the value which makes step 4 of the Proving Algorithm true, namely, set E(w)={<$v_i$, w, L-1>}∪ (Comp(E($v_i$), $v_i$, R(E)) of $G_2$). (We can execute step 1, 2, 3 at first so that we can get all Comp(E(x), x, R(E)) at stage L-2. More exactly, R(E) here should be R($E_2$ of $G_2$). We only use R(E) to represent all the reachable path sets of a graph, hence we just simply use R(E) in all cases in this paper. Actually readers know what E exactly is from the context. For example, '(Comp(E($v_i$), $v_i$, R(E)) of $G_2$)' implies that E means (E of $G_2$).)

Set $(/e_1/e_2/.../e_{|E_2|}$ of $G_2$)=(E of G) [2:2].

Set ($R_{ini}(u,v_1,L-2)$ of $G_2$)= $\underset{\{w|d(w)\,of\,stage\,L-1>1,\;after\,splitting\,v\}}{Y}$ ($I_w^{w_1,w_2}$ ($I_v^{v_1,v_2}$ (E of G, $group_1$, $group_2$), {<$v_1$, w, L-1>}, {<$v_2$, w, L-1>})); set ($R_{ini}(a,b,L-2)$ of $G_2$)=Ø if <a, b, L-2>≠<u, $v_1$, L-2>. (Which means, ($R_{ini}(u,v_1,L-2)$ of $G_2$) contains ($E_2$ of $G_2$), and only ($R_{ini}(u,v_1,L-2)$ of $G_2$) in all ($R_{ini}(a,b,L-2)$ of $G_2$) is not empty).

Set (ESS of $G_2$)= $I_v^{v_1,v_2}$ ((Comp(E(v), v, R(E)) of G) -{<$a_1$, $b_1$, k> | <$a_1$, $b_1$, k>≠<a, b, k>}, $group_1$, $group_2$)∪$E_2$[L-1：L]).

Thus we get a multistage graph $G_2$, as shown in Fig. 6(b). It is worthy noting that we only changed some edge sets of $G_1$ and we did not change the shape of $G_1$.

After applying the Proving Algorithm on $G_2$, we have the following three results (a), (b), (c).

(a) (Comp(ESS1, D, R(E)) of $G_2$)≠Ø.

(R(a, b, k) of G)(k<L-2) before step 5 contains <u, v, L-2>, hence $[R(a, b, k) \cap$ (Comp(E(v), v, R(E)) of G) $]_b^D$ contains <u, v, L-2>. (refer to R(a, b, k)[k+1：l] ← $\underset{v\in V_l}{U}$ [ R(a, b, k)∩Comp(E(v), v, R(E)) $]_b^v$ please). The iteration defined by step 2 and 3 will finally be stable. If (R(a, b, k) of G) contains <u, v, L-2>, according to the definition of Change(R(a, b, k)), (Comp([ {e | e=<c, d, kk>∈E, kk < k, [ R(e)∩Comp(E(v), v, R(E)) $]_d^v$ contains <u, v, L-2> and <a, b, k>} $]_S^a$, a, R(E)) of G)≠∅.

Since $E(v_1)=E(v_2)=I_v^{v_1,v_2}$ ((E of G)[1:L-2], group$_1$, group$_2$), and we have set E(w) with the value that makes step 4 true for all w at stage L-1, and (ESS of G$_2$)= $I_v^{v_1,v_2}$ ((Comp(E(v), v, R(E)) of G) -{<a$_1$, b$_1$, k> | <a$_1$, b$_1$, k>≠<a, b, k>}, group$_1$, group$_2$)∪E$_2$[L-1：L]), hence we know that, (Comp(E(D), D, R(E)) of G$_2$) in step 5 is not empty, and the expansion in step 7 is meaningless, and finally, (Comp(ESS1, D, R(E)) of G$_2$) in step 10 is not empty although <a, b, k> is the unique edge at stage k in ESS of G$_2$.

(b) ($/e_1/e_2/.../e_{|E_2|}$ of G$_2$) ∩ (Comp(ESS1, D, R(E)) of G$_2$) contains <aa, bb, 2>, such that [ R(aa, bb, 2) ∩ (Comp(ESS1, D, R(E)) of G$_2$) $]_{bb}^D$ contains bb– b$_3$ – b$_4$ – …– b$_{L-3}$– b$_{L-2}$– b$_{L-1}$–D, in which b$_{L-2}$– b$_{L-1}$– D is contained in ($R_{ini}(b_{L-3},b_{L-2},L-2)$ of G$_2$) and <b$_{L-3}$, b$_{L-2}$, L-2>=<u, v$_1$, L-2>.

The following facts result in (b): ($/e_1/e_2/.../e_{|E_2|}$ of G$_2$) = (E of G)[2:2]; ($R_{ini}(u,v_1,L-2)$ of G$_2$)= $\underset{\{w|d(w) of stage L-1>1,\ after splitting v\}}{Y}$ ($I_w^{w_1,w_2}$ ($I_v^{v_1,v_2}$ (E of G, group$_1$, group$_2$), {<v$_1$, w, L-1>}, {<v$_2$, w, L-1>})); $E(v_1)=E(v_2)=I_v^{v_1,v_2}$ ((E of G)[1:L-2], group$_1$, group$_2$); (ESS of G$_2$)= $I_v^{v_1,v_2}$ ((Comp(E(v), v, R(E)) of G) -{<a$_1$, b$_1$, k> | <a$_1$, b$_1$, k>≠<a, b, k>}, group$_1$, group$_2$)∪E$_2$[L-1：L]); and critically important, (Comp([ {e | e=<c, d, kk>∈E, kk < k, [ R(e)∩Comp(E(v), v, R(E)) $]_d^v$ contains <u, v, L-2> and <a, b, k>} $]_S^a$, a, R(E)) of G)≠∅.

(c) G$_2$<G.

The reason is simple. We did not change the shape of G$_1$ and we have proved G$_1$ < G, hence G$_2$ < G.

Based on (a), (b), (c), we can draw conclusion that there must exist in G$_2$ a simple path PP=S – a$_1$ – a$_2$ – …– a$_{L-3}$– a$_{L-2}$– a$_{L-1}$– D such that (ESS of G$_2$) contains PP, < a$_1$, a$_2$, 2> is in ($/e_1/e_2/.../e_{|E_2|}$ of G$_2$), and ($R_{ini}(a_{L-3},a_{L-2},L-2)$ of G$_2$) contains a$_{L-2}$– a$_{L-1}$– D. Since <a, b, k> is the unique edge at stage k in (ESS of G$_2$) and <u, v$_1$, L-2> is the unique edge at stage L-2 whose ($R_{ini}(u,v_1,L-2)$ of G$_2$) is not empty, both <a, b, k> and <u, v$_1$, L-2> must be on PP.

Thus we have proved that, if R(a, b, k) of G contains <u, v, L-2>, G must have a pre-simple path which traverses <a, b, k> and <u, v, L-2>, since $I_w^{w_2} I_w^{w_1}$ ($I_v^{v_2} I_v^{v_1}$ (ESS of G$_2$))[1:L-2] is a subset of (Comp(E(v), v, R(E)) of G).

(3) When $G_1$ is the input of the Proving Algorithm, we will have: (a) (Comp(ESS1, D, R(E)) of $G_1$)≠Ø, and (b) $/e_1/e_2/.../e_{|E_2|}$ ∩ (Comp(ESS1, D, R(E)) of $G_1$) contains <aa, bb, 2>, such that [ R(aa, bb, 2)∩(Comp(ESS1, D, R(E)) of $G_1$) $]_{bb}^{D}$ contains bb– $b_3$ – $b_4$ – …– $b_{L-3}$– $b_{L-2}$– $b_{L-1}$–D, in which $b_{L-2}$– $b_{L-1}$– D is contained in $R_{ini}(b_{L-3}, b_{L-2}, L-2)$.

We have not assigned values to all those vertices that are not D and originally on v– w– D before splitting so far. Using conclusion (2), we can assign values to $E(v_1)$, $E(v_2)$, $E(w_1)$, $E(w_2)$ now.

Set $E(v_1) = I_v^{v_1,v_2}$ ({e | e∈E(v), e is on a pre-simple path PP in E(v), PP[L-2:L-2] is in $group_1$}), $E(v_2) = I_v^{v_1,v_2}$ ({e | e∈E(v), e is on a pre-simple path PP in E(v), PP[L-2:L-2] is in $group_2$}).

For $v_1$–$w_1$–D in $G_1$, if (E(w) of G) ={<v, w, L-1>}∪ (Comp(E(v), v, R(E)) of G) and (Comp(E(v), v, R(E)) of G)[L-2:L-2]∩$group_1$ is not empty, set $E(w_1)$= {<$v_1$, $w_1$, L-1>}∪ (Comp(E($v_1$), $v_1$, R(E)) of $G_1$). For $v_2$–$w_2$–D in $G_1$, if (E(w) of G) ={<v, w, L-1>}∪ (Comp(E(v), v, R(E)) of G) and (Comp(E(v), v, R(E)) of G)[L-2:L-2]∩$group_2$ is not empty, set $E(w_2)$= {<$v_2$, $w_2$, L-1>}∪ (Comp(E($v_2$), $v_2$, R(E)) of $G_1$). Otherwise, set $E(w_1)$= $E(w_2)$ = $E_1$[L-2:L-2]( it means $E(w_1)$[L-1:L-1]= $E(w_2)$[L-1: L-1]=Ø and it gives an edge more chance to go through E(w) when its R(e) is computed).

Obviously, conclusion (2) here is the key basis that we can split v. If (R(a, b, k) of G) (k<L-2) contains <u, v, L-2> and <u, v, L-2> is in $group_1$, (R(a, b, k) of $G_1$) contains <u, $v_1$, L-2>. If (R(a, b, k) of G) (k<L-2) contains <u, v, L-2> and <u, v, L-2> is in $group_2$, (R(a, b, k) of $G_1$) contains <u, $v_2$, L-2>. This is the reason why we have $H_1 \wedge H_2$ after applying the Proving Algorithm on $G_1$.

A property which $E(v_1)$ and $E(v_2)$ hold is that both $I_v^{v_2} I_v^{v_1}$ ( $E(v_1)$) and $I_v^{v_2} I_v^{v_1}$ ( $E(v_2)$) are subset of E(v). Another property that $E(w_1)$ and $E(w_2)$ hold is that both $I_w^{w_2} I_w^{w_1} I_v^{v_2} I_v^{v_1}$ ( $E(w_1)$) and $I_w^{w_2} I_w^{w_1} I_v^{v_2} I_v^{v_1}$ ( $E(w_2)$) are subsets of E(w) if $E(w_1)$[L-1:L-1] and $E(w_2)$[L-1:L-1] are not empty. These two properties ensure that $I_w^{w_2} I_w^{w_1} I_v^{v_2} I_v^{v_1}$ (P) is a simple path in G if P is a simple path in $G_1$.

Now we check the following list. No matter whether it is before step 5 or after step 5, we have:

For all vertices x at stage k, k<L-2, (Comp(E(x), x, R(E)) of G) ⊆(Comp(E(x), x, R(E)) of $G_1$).

For all <a, b, k> (k<L-2) and for all <a, b, L-2> (b≠v), if (R(a, b, k) of G) contains a path b– $b_{k+1}$ – $b_{k+2}$ – …– $b_{L-3}$– $b_{L-2}$– $b_{L-1}$–D, (R(a, b, k) of $G_1$) contains b– $c_{k+1}$ – $c_{k+2}$ – …– $c_{L-3}$– $c_{L-2}$– $c_{L-1}$–D such that $I_w^{w_2} I_w^{w_1} I_v^{v_2} I_v^{v_1}$ (b– $c_{k+1}$ – $c_{k+2}$ – …– $c_{L-3}$– $c_{L-2}$– $c_{L-1}$–D)= b– $b_{k+1}$ – $b_{k+2}$ – …– $b_{L-3}$– $b_{L-2}$– $b_{L-1}$–D.

For all <u, v, L-2> in G, if (R(u, v, L-2) of G) contains v–w–D and <u, v, L-2>∈$group_1$, (R(u, $v_1$, L-2) of $G_1$) contains $v_1$–$w_1$–D; if (R(u, v, L-1) of G) contains v–w–D and <u, v, L-1>∈$group_2$, (R(u, $v_2$, L-1) of $G_1$) contains $v_2$–$w_2$–D.

Therefore, (ESS∩Comp(E(D), D, R(E))) of G)$\subseteq$ $I_w^{w_2} I_w^{w_1} I_v^{v_2} I_v^{v_1}$ (ESS∩Comp(E(D), D, R(E)) of $G_1$) in step 5, (Comp(ESS1, D, R(E)) of G)$\subseteq$ $I_w^{w_2} I_w^{w_1} I_v^{v_2} I_v^{v_1}$ (Comp(ESS1, D, R(E)) of $G_1$) in step 10, and after applying the Proving Algorithm on $G_1$, we have $H_1 \wedge H_2$ (that is, (Comp(ESS1, D, R(E)) of $G_1$)≠Ø; $/e_1/e_2/.../e_{|E_2|}$ ∩ (Comp(ESS1, D, R(E)) of $G_1$) contains <aa, bb, 2>, such that [ (R(aa, bb, 2) of $G_1$)∩(Comp(ESS1, D, R(E)) of $G_1$) $]_{bb}^{D}$ contains bb– $b_3$ – $b_4$ – …– $b_{L-3}$– $b_{L-2}$– $b_{L-1}$–D, in which $b_{L-2}$– $b_{L-1}$– D is contained in ($R_{ini}(b_{L-3}, b_{L-2}, L-2)$ of $G_1$).

(4) There exists a simple path P=S – $a_1$ – $a_2$ – …– $a_{L-3}$– $a_{L-2}$– $a_{L-1}$– D in G, such that ESS contains P, < $a_1$, $a_2$, 2> is in $/e_1/e_2/.../e_{|E_2|}$, and $R_{ini}(a_{L-3}, a_{L-2}, L-2)$ contains $a_{L-2}$– $a_{L-1}$– D.

We have assumed that G is the smallest graph that fails the Proving Algorithm and we have proved that $G_1$<G, hence there must exist a simple path P=S – $a_1$ – $a_2$ – …– $a_{L-3}$– $a_{L-2}$– $a_{L-1}$– D in $G_1$ such that (ESS of $G_1$) contains P, < $a_1$, $a_2$, 2> is in ($/e_1/e_2/.../e_{|E_2|}$ of $G_1$), and ($R_{ini}(a_{L-3}, a_{L-2}, L-2)$ of $G_1$) contains $a_{L-2}$– $a_{L-1}$– D.

If there exists a simple path P=S – $a_1$ – $a_2$ – …– $a_{L-3}$– $a_{L-2}$– $a_{L-1}$– D in $G_1$, such that (ESS of $G_1$) contains P, < $a_1$, $a_2$, 2> is in ($/e_1/e_2/.../e_{|E_2|}$ of $G_1$), and ($R_{ini}(a_{L-3}, a_{L-2}, L-2)$ of $G_1$) contains $a_{L-2}$– $a_{L-1}$– D, then, $I_w^{w_2} I_w^{w_1} I_v^{v_2} I_v^{v_1}$ (S – $a_1$ – $a_2$ – …– $a_{L-3}$– $a_{L-2}$– $a_{L-1}$– D) must be a simple path in G claimed in step 10. ∎

**Lemma 4.** Let G = <V, E, S, D, L> be the input of the Proving Algorithm, vertex v of stage $l$ is a multi in-degree vertex, $l$= L-3, and no multi in-degree vertex can be found at stage L-1 and L-2, as shown in Fig.7(a). After applying the Proving Algorithm on G, if we have

(1) Comp(ESS1, D, R(E))≠Ø,

(2) $/e_1/e_2/.../e_{|E_2|}$ ∩Comp(ESS1, D, R(E)) contains <aa, bb, 2>, such that [ R(aa, bb, 2)∩Comp(ESS1, D, R(E)) $]_{bb}^{D}$ contains bb– $b_3$ – $b_4$ – …– $b_{L-3}$– $b_{L-2}$– $b_{L-1}$–D, in which $b_{L-2}$– $b_{L-1}$– D is contained in $R_{ini}(b_{L-3}, b_{L-2}, L-2)$,

then, in G, there must exist a simple path P=S – $a_1$ – $a_2$ – …– $a_{L-3}$– $a_{L-2}$– $a_{L-1}$– D, such that ESS contains P, < $a_1$, $a_2$, 2> is in $/e_1/e_2/.../e_{|E_2|}$, and $R_{ini}(a_{L-3}, a_{L-2}, L-2)$ contains $a_{L-2}$– $a_{L-1}$– D.

**Proof:** We summarize the main idea of the proof firstly. To prove lemma 4, we need to construct a graph $G_1$ such that:

(1) $G_1$ is a multistage graph with L-1 stages;

(2) If we have $H_1 \wedge H_2$ when G is the input of the Proving Algorithm, we must have $H_1 \wedge H_2$ when $G_1$ is

the input of the Proving Algorithm.

(3) If P is a simple path in $G_1$, P must be a simple path in G.

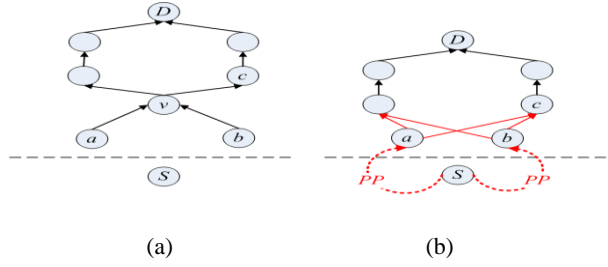

(a)        (b)

Figure 7 Typical graph of lemma 4

Proof begins. We can directly verify that the Proving Algorithm gives us correct claim for all multistage graphs with four stages. So, in the following discussion, we assume that the Proving Algorithm can give us correct claim for all multistage graphs with L-1 stages.

**If $l >2$.** We construct $G_1=<V_1, E_1, S, D, L-1>$ as follows (please note that $l=L-3$ in this case)**:**

(1) Vertices of $G_1$: $V_1=$ (V of G)- ($V_{L-3}$ of G). (Please noting: $G_1$ has no vertex at stage L-3 of G.)

(2) Edges of $G_1$**:** For all $<g_1, g_2, k>$ in G, if $k<l$, $<g_1, g_2, k>$ becomes an edge in $G_1$, if $k > l+1$, $< g_1, g_2, k-1>$ becomes an edge in $G_1$. For all $<a, v, l>$ and $<v, c, l+1>$ in G, $v\in$ (V of G), we generate $<a, c, l>$ in $G_1$.

(3) Edge sets of $G_1$ **:**

For all $x\in V_1$, if x is not at stage $l$($l=L-3=L-1-2$ in this case), set (E(x) of $G_1$)=((E(x) of G)-{e | e∈(E of G), e ends at v or starts from v, v∈ ($V_l$ of G)})∪{e | e=$<a, c, l>$ , there exists v such that $<a, v, l>$ and $<v, c, l+1>$ ∈(E(x) of G)}.

For all $c\in V_1$, if c is just at stage $l$, set (E(c) of $G_1$)= ((Comp(E(c), c, R(E)) of G) −{e | e∈(E of G), and e ends at v or starts from v, v∈ ($V_l$ of G) })∪{e | e=$<a, c, l>$, there exists v such that $<a, v, l>$, $<v, c, l+1>$ ∈(Comp(E(c), c, R(E)) of G)}. The result of (Comp(E(c), c, R(E)) of G) is variable at different steps. Here we choose the value before the execution of step 5.

(4) ESS of $G_1$:

Set (ESS of $G_1$)=(( ESS of G)-{e | e∈(E of G), e ends at v or starts from v, v∈ ($V_l$ of G) })∪{e | e=$<a, c, l>$, there exists v such that $<a, v, l>$ and $<v, c, l+1>$∈(ESS of G)}

(5) $/e_1/e_2/.../e_{|E_2|}$ of $G_1$:

Set $(/e_1/e_2/.../e_{|E_2|}$ of $G_1)=(/e_1/e_2/.../e_{|E_2|}$ of G).

For all $<a, c, l>$ in $G_1$, add a path PP from S to vertex a in $G_1$, expand all edge sets (including (ESS of $G_1$) and edge sets of the vertices on PP, except ($/e_1/e_2/.../e_{|E_2|}$ of $G_1$)) to include the edges on PP.

For all $<f, g, k>$ in $G_1$, $k<l$, add a path PP from S to vertex f in $G_1$, expand all edge sets (including

(ESS of $G_1$) and edge sets of the vertices on PP, except ($/e_1/e_2/.../e_{|E_2|}$ of $G_1$)) to include the edges on PP.

(Please note: If $<c, w, l+1>$ is in $G_1$ and $E(w)[l+1:l+1]$ is not empty, $E(w)$ has been expanded to include the edges on PP. Therefore, step 4 will not set all R(e) empty when we apply the Proving Algorithm on $G_1$.)

(6) $R_{ini}(a,b,L-1-2)$ of $G_1$:

If $[R(aa, bb, 2) \cap (Comp(ESS1, D, R(E))$ of $G)]_{bb}^D$ contains $bb - b_3 - b_4 - ... - b_{L-3} - b_{L-2} - b_{L-1} - D$, in which $b_{L-2} - b_{L-1} - D$ is contained in ($R_{ini}(b_{L-3}, b_{L-2}, L-2)$ of G), then, we set ($R_{ini}(cc_{L-4}, cc_{L-3}, L-1-2)$ of $G_1$)={e | e is on $b_{L-2} - b_{L-1} - D$}; we set all ($R_{ini}(a,b,L-1-2)$ of $G_1$) =Ø if $<a, b, L-1-2> \neq <cc_{L-4}, cc_{L-3}, L-1-2>$, where, $cc_{L-4} = b_{L-4}$, $cc_{L-3}=b_{L-2}$, $<a, b, L-1-2> \in E_1$.

(7) Modification of all edge sets and ESS:

For all $<g_1, g_2, k>$ in G, for all (E(x) of $G_1$) that contains $<g_1, g_2, k>$, if $<g_1, g_2, k>$ in G has become $<g_1, g_2, k-1>$ in $G_1$, use $<g_1, g_2, k-1>$ to substitute $<g_1, g_2, k>$ in (E(x) of $G_1$).

(8) Expanding (E(c) of $G_1$) at stage $l$ ($l$=L-3=L-1-2 in this case):

For all $c \in V_1$, if c is just at stage $l$, excute the following 3 steps:

Expand (E(c) of $G_1$) to include all edges ending at c in $G_1$

For all $w \in V_1$, if $<c, w, l+1> \in E_1$, set (E(w) of $G_1$)[1:$l$]= (E(c) of $G_1$)

Expand (E(c) of $G_1$) to include all edges at stage 1 and stage 2.

The constructed $G_1$ is shown in Fig.7(b).

To help us understand the proof, some key points are explained at first.

(1) Why do we set

(E(c) of $G_1$)= ((Comp(E(c), c, R(E)) of G) −{e | e∈(E of G), and e ends at v or starts from v, $v \in V_l$ of G }) ∪ {e | e=$<a, c, l>$, there exists v such that $<a, v, l>$, $<v, c, l+1> \in$ (Comp(E(c), c, R(E)) of G)},

and choose the value of (Comp(E(c), c, R(E)) of G ) before step 5 to set E(c)?

(Comp(E(c), c, R(E)) of G) before step 5 is a subset of the initial (E(c) of G). (Comp(E(c), c, R(E)) of G)[1:L-3] is also a subset of the initial (E(v) of G). Therefore, if (E(c) of $G_1$) contains a path P, initial E(c) and E(v) in G contain P[1:L-3].

(2) Why do we add PP in $G_1$?

Firstly, <a, c, *l*> may not be kept in (Comp(E(c), c, R(E)) of $G_1$) without PP.

Secondly, there may be some problems in computing Change(R(f, g, k)) (k<*l*) without PP. In G, when Change(R(f, g, k)) decides that <v, c, *l*+1> is remained in R(f, g, k), edges on a path from S to f will tranverse <v, c, *l*+1> with <f, g , k>. Maybe some edges on the path tranverse <a, v, *l*> and <v, c, *l*+1> while others tranverse <b, v, *l*> and <v, c, *l*+1>. However, in $G_1$, all the edges on the path are forced to tranverse <a, c, *l*> or <b, c, *l*>. Therefore, <a, c, *l*> or <b, c, *l*> get no guarantee to be kept in R(f, g, k)(k<*l*). This is the reason why we add PP in $G_1$. After we add PP in $G_1$, <a, c, *l*> or <b, c, *l*> can be kept in (R(f, g, k) of $G_1$).

(3) Why do not we expand $/e_1/e_2/.../e_{|E_2|}$ to include PP[2:2]?

If we expand $/e_1/e_2/.../e_{|E_2|}$ to include PP[2:2], a simple path which contains PP[2:2] in $G_1$ will not be able to become a simple path in G.

(4) Why do we in the Proving Algorithm expand all edge sets of vertices at stage L-1 and L-2 and repeat step 1, 2, 3 again?

The expansion of step 7 in the Proving Algorithm plays an important role to hold $H_1 \wedge H_2$ for Lemma 4.

In fact, after execution of R(a, b, k)[k+1: L-2] ← [ R(a, b, k)∩ $A_1$ ]$_b^{w_1}$ ∪ … ∪ [ R(a, b, k)∩ $A_j$ ]$_b^{w_j}$ , if (R(a, b, k) of G) (when G is the input of the Proving Algorithm) gets a path from b to D, then, (R(a, b, k) of $G_1$) (when $G_1$ is the input of the Proving Algorithm) can get a path from b to D too. However, if (R(a, b, k) of G) gets two paths, i.e., b– $b_{k+1}$ – $b_{k+2}$ – …– $b_{L-3}$–v–…–w–D and b– $c_{k+1}$ – $c_{k+2}$ – …– $c_{L-3}$–v–…–$w_1$–D, (R(a, b, k) of G) may form b– $b_{k+1}$ – $b_{k+2}$ – …– $b_{L-3}$–v–…–$w_1$–D and b– $c_{k+1}$ – $c_{k+2}$ – …– $c_{L-3}$–v–…–w–D. Even if (ESS of G) does not contain v–…–$w_1$–D and v–…–w–D at the same time, <a, b, k> may be contained in (Comp(ESS, D, R(E) of G). In this case, however, R(a, b, k) in $G_1$ may contain two paths and may not form a new path, therefore, <a, b, k> may have no chance to be contained in (Comp(ESS, D, R(E) of $G_1$). After expanding all the said edge sets and executing step 1,2,3 again, we meet no problem of this kind again.

(5) Why do we expand (E(c) of $G_1$) at stage *l* (*l*=L-3=L-1-2 in this case)?

Firstly, after expanding (E(c) of $G_1$) to include all edges ending at c in $G_1$, all edges ending at c can be kept in (Comp(E(D), D, R(E)) of $G_1$) (refer to step 5 please).

Secondly, after expanding (E(c) of $G_1$) to include all edges at stage 1 and stage 2, we can have $H_2$ (refer to step 10 please) because the Proving Algorithm do not expand edge sets at stage L-2 to include E[1:2] (refer to step 7 please).

Thirdly, after expanding (E(c) of $G_1$) to include all edges ending at c and all edges at stage 1 and stage 2, (E(c) of $G_1$) may contain pre-simple path P such that P[1:2]∈ {e| e∈$E_1$, e is at stage 1 or stage 2, e is not in (Comp(E(c), c, R(E)) of G) , e is not on PP } (PP is the path we add in $G_1$). Hence for all w∈$V_1$, we set (E(w) of $G_1$)[1:*l*]= (E(c) of $G_1$) if <c, w, *l*+1>∈$E_1$.

After setting (E(w) of $G_1$)[1:*l*]= (E(c) of $G_1$), no edges in {e| e∈$E_1$, e is at stage 1 or stage 2, e is not in

(Comp(E(c), c, R(E)) of G) , e is not on PP } is on a simple path which traverses vertex c in $G_1$.

After setting (E(w) of $G_1$)[1:$l$]= (E(c) of $G_1$), (Comp(E(c) of $G_1$, c, R(E)) of $G_1$) contains no edges in {e| e ∈$E_1$, e is at stage 1 or stage 2, e is not in (Comp(E(c), c, R(E)) of G) , e is not on PP }. Step 4 is true.

Go back to the proof. We can prove the following conclusion (1), (2), (3).

(1) After applying the Proving Algorithm on $G_1$, ( /$e_1$ /$e_2$ /…/ $e_{|E_2|}$ of $G_1$)∩ (Comp(ESS1, D, R(E)) of $G_1$) will contain <aa, bb, 2>, such that [ R(aa, bb, 2)∩ (Comp(ESS1, D, R(E)) of $G_1$) ]$_{bb}^{D}$ contains bb– $b_3$ – $b_4$ – …– $b_{L-4}$– $b_{L-2}$– $b_{L-1}$–D, in which $b_{L-2}$– $b_{L-1}$– D is contained in ( $R_{ini}(b_{L-4}, b_{L-2}, L-1-2)$ of $G_1$).

The reason is as follows. Before step 5, if (Comp(E(D), D, R(E)) of G) contains <a, b, k> (k<$l$ or k>$l$+1), (Comp(E(D), D, R(E)) of $G_1$) must contain <a, b, k>. If (Comp(E(D), D, R(E)) of G) contains <v, c, $l$+1>, there must exist an edge ending at c in (Comp(E(D), D, R(E)) of $G_1$), since (Comp(E(D), D, R(E)) of G) ⊇ (Comp(E(c), c, R(E)) of G) and (Comp(E(D), D, R(E)) of $G_1$) ⊇ (Comp(E(c), c, R(E)) of $G_1$).

After step 5, we expand all edge sets at stage L-1 and L-2 in G, and expand all edge sets at L-1-1 and L-1-2 in $G_1$. If (Comp(ESS1, D, R(E)) of G) contains <a, v, $l$ > and <v, c, $l$+1>, there must exist <a, c, $l$ > in (Comp(ESS1, D, R(E)) of $G_1$) and (Comp(E(c), c, R(E)) of $G_1$).

Noting the fact that ( /$e_1$ /$e_2$ /…/ $e_{|E_2|}$ of G)∩ (Comp(ESS1, D, R(E)) of G) contains <aa, bb, 2>, such that [ R(aa, bb, 2)∩ (Comp(ESS1, D, R(E)) of G) ]$_{bb}^{D}$ contains bb– $b_3$ – $b_4$ – …– $b_{L-3}$– $b_{L-2}$– $b_{L-1}$–D, in which $b_{L-2}$– $b_{L-1}$–D is contained in $R_{ini}(b_{L-3}, b_{L-2}, L-2)$, we can draw the conclusion that ( /$e_1$ /$e_2$ /…/ $e_{|E_2|}$ of $G_1$) ∩ (Comp(ESS1, D, R(E)) of $G_1$) contain <aa, bb, 2>, such that [ R(aa, bb, 2)∩ (Comp(ESS1, D, R(E)) of $G_1$) ]$_{bb}^{D}$ contains bb– $b_3$ – $b_4$ – …– $b_{L-4}$– $b_{L-2}$– $b_{L-1}$–D, in which $b_{L-2}$– $b_{L-1}$– D is contained in $R_{ini}(b_{L-4}, b_{L-2}, L-1-2)$, (Here, <$b_{L-4}$, $b_{L-3}$, L-3> and <$b_{L-3}$, $b_{L-2}$, L-2> in G is substituted by <$b_{L-4}$,$b_{L-2}$, L-1-2> in $G_1$).

(2) There must exist a simple path P=S – $a_1$ – $a_2$ – …– $a_{L-4}$– $a_{L-2}$– $a_{L-1}$– D in $G_1$ such that (ESS of $G_1$) contains P, < $a_1$, $a_2$, 2> is in /$e_1$ /$e_2$ /…/ $e_{|E_2|}$, and $R_{ini}(a_{L-4}, a_{L-2}, L-2)$ contains $a_{L-2}$– $a_{L-1}$– D.

The reason is very simple. The stage number of $G_1$ is L-1.

(3) If there exists a simple path P=S – $a_1$ – $a_2$ – …– **$a_{L-4}$**– **$a_{L-2}$**– **$a_{L-1}$**– **D** in $G_1$ (Please note: P does not contain **$a_{L-3}$**) such that (ESS of $G_1$) contains P, < $a_1$, $a_2$, 2> is in ( /$e_1$ /$e_2$ /…/ $e_{|E_2|}$ of $G_1$), and $R_{ini}(a_{L-4}, a_{L-2}, L-1-2)$ contains $a_{L-2}$– $a_{L-1}$– D, then, PPP=S – $a_1$ – $a_2$ – …– **$a_{L-4}$**– **$a_{L-3}$**– **$a_{L-2}$**– **$a_{L-1}$**– **D** is certainly to be a simple path in G (Please note: PPP contains **$a_{L-3}$** here), (ESS of G) contains PPP, ( /$e_1$ /$e_2$ /…/ $e_{|E_2|}$ of G) contains <$a_1$, $a_2$, 2>, and ($R_{ini}(a_{L-3}, a_{L-2}, L-2)$ of G) contains $a_{L-2}$– $a_{L-1}$– D.

The reason is as follows:

(ESS of $G_1$)=(( ESS of G)-{e | e∈(E of G), e ends at v and starts from v })∪{e | e=<a, c, $l$>, there exists v such that <a, v, $l$> and <v, c, $l$+1>∈(ESS of G)}. (ESS of $G_1$) contains P implies that (ESS of G) contains PPP.

(E($a_{L-2}$) of $G_1$)= ((Comp(E($a_{L-2}$), $a_{L-2}$, R(E)) of G) –{e | e∈(E of G), and e ends at v or starts from v})∪{e | e=<a, c, (L-1)-2>, there exists v such that <a, v, $l$>, <v, c, $l$+1> ∈(Comp(E($a_{L-2}$), $a_{L-2}$, R(E)) of G)}. (E($a_{L-2}$) of $G_1$) contains P[1: (L-1)-2] implies that (E($a_{L-2}$) of G) contains PPP[1:L-3] and (E($a_{L-3}$) of G) contains PPP[1:L-3].

**If $l$ = 2,** as shown in Fig.8(a) (please note that $l$=L-3 in this case).

We can directly prove that the Proving Algorithm makes correct assertion in this case.

We have assumed that, Comp(ESS1, D, R(E))≠Ø, and $/e_1/e_2/.../e_{|E_2|}$ ∩Comp(ESS1, D, R(E)) contains <aa, bb, 2>, such that [ R(aa, bb, 2)∩Comp(ESS1, D, R) $]_{bb}^{D}$ contains bb– $b_3$ – $b_4$ – …– $b_{L-3}$– $b_{L-2}$– $b_{L-1}$–D, in which $b_{L-2}$– $b_{L-1}$– D is contained in $R_{ini}(b_{L-3}, b_{L-2}, L-2)$, hence we know that there must be an edge <a, v, 2> such that R(a, v, 2)∩Comp(ESS1, D, R(E)) contains v – c – w – D. Noting the facts that E(w)={<c, w, L-1>}∪Comp(E(c), c, R(E)) in which Comp(E(c), c, R(E)) is the value before step 5, and we do not expand E(c) to include E[1:2](refer to step 7 please), S – a – v – c – w – D must be a simple path in G. (Here we need a patch and each Lemma should have this patch )

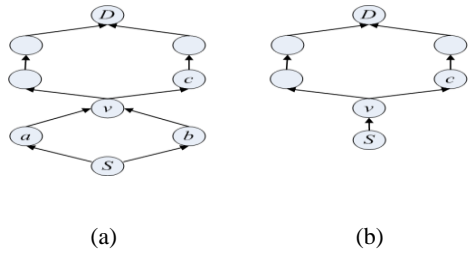

(a)            (b)

Figure 8 Typical graph of lemma 4 when $l \leq 2$

**If $l$ = 1,** as shown in Fig.8(b), we can similarly prove that the proving algorithm makes correct assertion in this case. ∎

**Lemma 5.** Let G = <V, E, S, D, L> be the input of the Proving Algorithm, vertex v of stage $l$ is a multi in-degree vertex, $l$ < L-3, and no multi in-degree vertex can be found at stage L-1, stage L-2,…, stage $l$+1, as shown in Fig. 9(a). After applying the Proving Algorithm on G, if we have

(1) Comp(ESS1, D, R(E))≠Ø,

(2) $/e_1/e_2/.../e_{|E_2|}$ ∩Comp(ESS1, D, R(E)) contains <aa, bb, 2>, such that [ R(aa, bb, 2)∩Comp(ESS1, D, R(E)) $]_{bb}^{D}$ contains bb– $b_3$ – $b_4$ – …– $b_{L-3}$– $b_{L-2}$– $b_{L-1}$–D, in which $b_{L-2}$– $b_{L-1}$– D is contained in $R_{ini}(b_{L-3}, b_{L-2}, L-2)$,

then, in G, there must exist a simple path P=S − $a_1$ − $a_2$ − …− $a_{L-3}$− $a_{L-2}$− $a_{L-1}$− D such that ESS contains P, < $a_1$, $a_2$, 2> is in $/e_1/e_2/.../e_{|E_2|}$, and $R_{ini}(a_{L-3}, a_{L-2}, L-2)$ contains $a_{L-2}$− $a_{L-1}$− D.

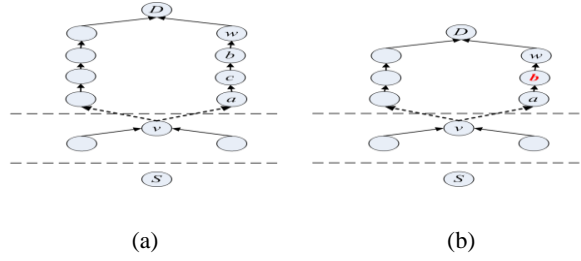

(a)    (b)

Figure 9 Typical graph of lemma 5

**Proof:** We summarize the main idea of the proof firstly. To prove Lemma 5, we need to construct a graph $G_1$ such that:

(1) $G_1$ is a multistage graph with L-1 stages;

(2) If we have $H_1 \wedge H_2$ when G is the input of the Proving Algorithm, we must have $H_1 \wedge H_2$ when $G_1$ is the input of the Proving Algorithm.

(3) If P is a simple path in $G_1$, P must be a simple path in G.

Proof begins. We can directly verify that the Proving Algorithm gives us correct claim for all multistage graphs with four stages. So, in the following discussion, we assume that the Proving Algorithm can give us correct claim for all multistage graphs with L-1 stages.

We construct $G_1$=<$V_1$, $E_1$, S, D, L-1> as follows (please note that $l$<L-3 in this case):

(1) Vertices of $G_1$: $V_1$= (V of G)- ($V_{L-3}$ of G). (Please noting: We have no vertex at stage L-3 of G.)

(2) Edges of $G_1$: For all <$g_1$, $g_2$, k> in G, if k<L-3, <$g_1$, $g_2$, k> becomes an edge in $G_1$, if k > L-3+1, <$g_1$, $g_2$, k-1> becomes an edge in $G_1$. For any <a, c, L-3> and <c, b, L-2> in G, we generate <a, b, (L-1)-2> in $G_1$.

(3) Edge sets of $G_1$:

For all x∈$V_1$, if x is not at stage L-1-2, set (E(x) of $G_1$)=((E(x) of G)-{e | e∈(E of G), e ends at c or starts from c, c is a vertex at stage L-3 in G })∪{e | e=<a, b, (L-1)-2>, there exists c such that <a, c, L-3> and <c, b, L-2>∈(E(x) of G)}.

For all b∈$V_1$, if b is just at stage L-1-2, set (E(b) of $G_1$)= ((Comp(E(b), b, R(E)) of G) −{e | e∈ (Comp(E(b), b, R(E)) of G), and e ends at c or starts from c, c is a vertex at stage L-3 in G })∪{e | e=<a, b, (L-1)-2>, there exists c such that <a, c, L-3>, <c, b, L-2> ∈(Comp(E(c), c, R(E)) of G)}. We have several values for (Comp(E(c), c, R(E)) of G) at different steps. Here we choose the value before the execution of step 5.

(4) ESS of $G_1$:

(ESS of $G_1$)=(( ESS of G)-{e | e∈(E of G), e ends at c or starts from c, c is a vertex at stage L-3 in G })∪{e | e=<a, b, (L-1)-2>, there exists c such that <a, c, L-3> and <c, b, L-2>∈(ESS of G)}.

(5) $R_{ini}(a,b,L-1-2)$:

If $[R(aa, bb, 2) \cap (Comp(ESS1, D, R(E))$ of $G)]_{bb}^{D}$ contains $bb - b_3 - b_4 - \ldots - b_{L-3} - b_{L-2} - b_{L-1} - D$, in which $b_{L-2} - b_{L-1} - D$ is contained in ($R_{ini}(b_{L-3},b_{L-2},L-2)$ of G), then, we set ($R_{ini}(cc_{L-4}, cc_{L-3}, L-1-2)$ of $G_1$)={e | e is on $b_{L-2} - b_{L-1} - D$}; set all ($R_{ini}(a,b,L-1-2)$ of $G_1$)=Ø if <a, b, L-1-2>≠< $cc_{L-4}$, $cc_{L-3}$, L-1-2> , where $cc_{L-4} = b_{L-4}$, $cc_{L-3}=b_{L-2}$, $b_{L-3}$ is not in $G_1$, <a, b, L-1-2>∈$E_1$.

(6) modification of all edge sets and ESS:

For all <$g_1$, $g_2$, k> in G, for all (E(x) of $G_1$) that contains <$g_1$, $g_2$, k>, if <$g_1$, $g_2$, k> in G has become <$g_1$, $g_2$, k-1> in $G_1$, use <$g_1$, $g_2$, k-1> to substitute <$g_1$, $g_2$, k> in (E(x) of $G_1$).

The constructed $G_1$ is shown as Fig. 9(b).

The following conclusions are obvious.

(1) After applying the Proving Algorithm on $G_1$, ($/e_1/e_2/\ldots/e_{|E_2|}$ of $G_1$)∩ (Comp(ESS1, D, R(E)) of $G_1$) will contain <aa, bb, 2>, such that $[R(aa, bb, 2) \cap (Comp(ESS1, D, R(E))$ of $G_1)]_{bb}^{D}$ contains $bb - b_3 - b_4 - \ldots - b_{L-4} - b_{L-2} - b_{L-1} - D$, in which $b_{L-2} - b_{L-1} - D$ is contained in ($R_{ini}(b_{L-4}, b_{L-2}, L-1-2)$ of $G_1$) (here, < $b_{L-4}$, $b_{L-3}$, L-3> and < $b_{L-3}$, $b_{L-2}$, L-2> in G are substituted by < $b_{L-4}$, $b_{L-2}$, L-1-2> in $G_1$).

It is a direct result from the assumption of Lemma5, i.e., ($/e_1/e_2/\ldots/e_{|E_2|}$ of G)∩ (Comp(ESS1, D, R(E)) of G) contains <aa, bb, 2>, such that $[R(aa, bb, 2) \cap (Comp(ESS1, D, R(E))$ of $G)]_{bb}^{D}$ contains $bb - b_3 - b_4 - \ldots - b_{L-3} - b_{L-2} - b_{L-1} - D$, in which $b_{L-2} - b_{L-1} - D$ is contained in ($R_{ini}(b_{L-3}, b_{L-2}, L-2)$ of G).

(2) There must exists a simple path P=S − $a_1$ − $a_2$ − …− $a_{L-4}$ − $a_{L-2}$ − $a_{L-1}$ − D (Please note: P does not contain **$a_{L-3}$**) in $G_1$ such that (ESS of $G_1$) contains P, <$a_1$, $a_2$, 2> is in ($/e_1/e_2/\ldots/e_{|E_2|}$ of $G_1$), and ($R_{ini}(a_{L-4}, a_{L-2}, L-2)$ of $G_1$) contains $a_{L-2} - a_{L-1} - D$.

The reason is that the stage number of $G_1$ is L-1.

(3) If there exists a simple path P=S − $a_1$ − $a_2$ − …− $a_{L-4}$ − $a_{L-2}$ − $a_{L-1}$ − D in $G_1$ (Please note: P does not contain **$a_{L-3}$**), such that (ESS of $G_1$) contains P, ($/e_1/e_2/\ldots/e_{|E_2|}$ of $G_1$)∩ (Comp(ESS1, D, R(E)) of $G_1$) contains <$a_1$, $a_2$, 2>, and ($R_{ini}(a_{L-4}, a_{L-2}, L-1-2)$ of $G_1$) contains $a_{L-2} - a_{L-1} - D$, then, PPP=S − $a_1$ − $a_2$ − …− $a_{L-4}$ − **$a_{L-3}$** − $a_{L-2}$ − $a_{L-1}$ − D is certainly a simple path in G (Please note: PPP contains the vertex $a_{L-3}$), (ESS of G) contains PPP, $(/e_1/e_2/.../e_{|E_2|}$ of G)∩ (Comp(ESS1, D, R(E)) of G) contains <$a_1$, $a_2$, 2>, and ($R_{ini}(a_{L-3}, a_{L-2}, L-2)$ of G) contains $a_{L-2}$– $a_{L-1}$– D.

The reason is as follows:

(ESS of $G_1$)=(( ESS of G)-{e | e∈(E of G), e ends at c or starts from c, c is a vertex at stage L-3 in G })∪{e | e=<a, b, (L-1)-2>, there exists c such that <a, c, L-3> and <c, b, L-2>∈(ESS of G)}. (ESS of $G_1$) contains P implies that (ESS of G) contains PPP.

(E($a_{L-2}$) of $G_1$)= ((Comp(E($a_{L-2}$), $a_{L-2}$, R(E)) of G) –{e | e∈(Comp(E($a_{L-2}$), $a_{L-2}$, R(E)) of G), e ends at c or starts from c, c is a vertex at stage L-3 in G})∪{e | e=<a, b, (L-1)-2>, there exists c such that <a, c, L-3>, <c, b, L-2> ∈(Comp(E(c), c, R(E)) of G)}. (E($a_{L-2}$) of $G_1$) contains P[1: (L-1)-2] implies that (E($a_{L-2}$) of G) contains PPP[1:L-2] and (E($a_{L-3}$) of G) contains PPP[1:L-3]. ∎

**αβ lemma**  Let G = <V, E, S, D, L> be a multistage graph. After applying the Proving Algorithm on G, if we have

(1) Comp(ESS1, D, R(E))≠Ø,

(2) $/e_1/e_2/.../e_{|E_2|}$ ∩Comp(ESS1, D, R(E)) contains <aa, bb, 2>, such that [ R(aa, bb, 2)∩Comp(ESS1, D, R(E)) $]_{bb}^{D}$ contains bb– $b_3$ – $b_4$ – …– $b_{L-3}$– $b_{L-2}$– $b_{L-1}$–D, in which $b_{L-2}$– $b_{L-1}$– D is contained in $R_{ini}(b_{L-3}, b_{L-2}, L-2)$,

then, in G, there must exist a simple path P=S – $a_1$ – $a_2$ – …– $a_{L-3}$– $a_{L-2}$– $a_{L-1}$– D, such that ESS contains P, <$a_1$, $a_2$, 2> is in $/e_1/e_2/.../e_{|E_2|}$, and $R_{ini}(a_{L-3}, a_{L-2}, L-2)$ contains $a_{L-2}$– $a_{L-1}$– D.

**Proof:**   We can verify directly that **αβ lemma** is true when L=4. We assume that **αβ lemma** can hold for all multistage graphs with L-1 stages**.**

If **αβ lemma** is false for some multistage graphs with L stages, we can find out the smallest graph that makes **αβ lemma** fail using the linear order "⩽" defined above. So, without losing generality, we can assume that G is the smallest graph which makes the proving algorithm fail when determining the existence of a simple path in G. **We prove the non-existence of this smallest graph.**

From all the discussion above, we know that**:**

(1) G must have a multi in-degree vertex since all graphs of lemma 1 will make **αβ lemma** true.

(2) According to lemma 2, G can not have a multi in-degree vertex at stage L-1.

(3) According to lemma 3, G can not have a multi in-degree vertex at stage L-2.

(4) According to lemma 4, G can not have a multi in-degree vertex at stage L-3.

(5) According to lemma 5, G can not have a multi in-degree vertex at stage $l$, $l$<L-3.

Therefore, no such smallest graph which makes **αβ lemma** incorrect can be found. ∎

## 3.4  Proving the sufficiency of   Z-H algorithm

We now return to **Z-H algorithm**. Using **αβ lemma**, we prove the existence of a simple path in G if Comp(E(D), D, R(E))≠Ø.

**Theorem 3** Let G=<V, E, S, D, L> be a multistage graph. After applying Z-H algorithm on G, if Comp(E(D), D, R(E))≠Ø, there must exist a simple path in G.

**Proof.** Based on G (shown in Fig.10(a)), we construct a multistage graph $G_{\alpha\beta}$. The stage number of $G_{\alpha\beta}$ is L+4.

(1) Add $S_0$－a－S in G, change all <a, b, k> in G into <a, b, k+2>, expand all edge sets of G to include < $S_0$, a, 1> and < a, S, 2>, set E(S)={< $S_0$, a, 1>, < a, S, 2>}, E(a)={< $S_0$, a, 1>}, set $/e_1/e_2/.../e_{|E_2|}$={<a, S, 2>}.

(2) Add D－w－$D_{\alpha\beta}$ in G, set $E(w)$ =E($D_{\alpha\beta}$)=(Comp(E(D), D, R(E)) of G)∪{ <D, w, L+3>, <w, $D_{\alpha\beta}$, L+4> }.

(3) For all vertices a at stage L+4-3 in $G_{\alpha\beta}$, set $R_{ini}(a,D,L+4-2)$={<D, w, L+3>, <w, $D_{\alpha\beta}$, L+4}.

(4) Set ESS = E($D_{\alpha\beta}$).

Thus we get a new multistage graph $G_{\alpha\beta}$=<$V_{\alpha\beta}$, $E_{\alpha\beta}$, $S_0$, $D_{\alpha\beta}$, L+4>, as shown in Fig.10(b).

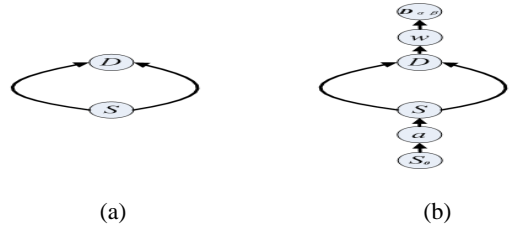

Figure 10 Relations between G and $G_{\alpha\beta}$

Now we apply the Proving Algorithm on $G_{\alpha\beta}$.

According to the construction of $G_{\alpha\beta}$, each edge set E(v) contains $S_0$－a－S, so, (Comp(E(v), v, R(E)) of $G_{\alpha\beta}$)=(Comp(E(v), v, R(E)) of G)∪{< $S_0$, a, 1>, < a, S, 2>}.

According to the construction of $G_{\alpha\beta}$, ESS=E(w)=E($D_{\alpha\beta}$)=(Comp(E(D), D, R(E)) of G)∪{ <D, w, L+3>, <w, $D_{\alpha\beta}$, L+4> }, so, when we apply the proving algorithm on $G_{\alpha\beta}$, step 7 becomes useless.

We have assumed that (Comp(E(D), D, R(E)) of G) ≠Ø after we applying **Z-H algorithm** on G, hence we know, after applying the proving algorithm on $G_{\alpha\beta}$, we will have

(1) (Comp(ESS1, $D_{\alpha\beta}$, R(E)) of $G_{\alpha\beta}$)≠Ø,

(2) $/e_1/e_2/.../e_{|E_2|}$ ∩ (Comp(ESS1, $D_{\alpha\beta}$, R(E)) of $G_{\alpha\beta}$) contains < a, S, 2>, such that [ R(a, S, 2)∩ (Comp(ESS1, $D_{\alpha\beta}$, R(E)) of $G_{\alpha\beta}$) $]_S^{D_{\alpha\beta}}$ contains S – $b_1$ – $b_2$ – …– $b_{L-3}$– $b_{L-2}$– $b_{L-1}$–D–w–$D_{\alpha\beta}$, in which D–w–$D_{\alpha\beta}$ is contained in $R_{ini}(b_{L-1},D,L+2)$.

Thus we know that there must exist a simple path in $G_{\alpha\beta}$ by **αβ lemma.** Assuming that $S_0 - a - S - v_1 - v_2 - \ldots - v_{L-1} - D - w - D_{\alpha\beta}$ is a simple path in $G_{\alpha\beta}$, $S - v_1 - v_2 - \ldots - v_{L-1} - D$ must be a simple path in G. ∎

## 4.  Proving MSP∈NPC

We now concentrate ourselves on determining the Hamilton property of an undirected graph. For a given undirected graph G=<V, E> of order n, we transform it into a labeled multistage graph G' = <V', E', S, D, L> according to the following six steps**:**

(1) Let L=n
(2) Select vertex v from V. Generate vertices (v, 0), (v, n) in V'. Let S＝(v, 0) and D＝(v, n)
(3) For all u∈V- {v}, generate (u, 1) of stage l, (u, 2) of stage 2, …, (u, L-1) of stage L-1 in V'
(4) For all (a, b)∈E, a≠v and b≠v, generate edges from previous stage to next stage in E'. These edges are <(a,1), (b,2), 2>,…,<(a,L-2), (b,L-1),L-1> and <(b,1), (a,2), 2>, …,<(b,L-2), (a,L-1), L-1>.
For all (v, b)∈E, generate edges from (v, 0) to (b, 1) and generate edges from (b, L-1) to (v, n) in E'.
(5) Let E(u, $l$) = E'－{e | e∈E' and e is associated with (u,1), …,or (u, $l$-1)}, 1≤$l$≤L-1.
(6) Let E(D)=E(v, n)=E'

Step 5 is important. The edge set of (u, $l$) is assigned with a value that permits the appearance of u in stage $l$ and forbids the appearance of u in those stages smaller than $l$.

**Theorem 4**  G is a Hamilton graph if and only if G' has a simple path from S to D.

**Proof**  If $v - a_1 - a_2 - \ldots - a_{n-1} - v$ is a Hamilton circle of G, $(v,0) - (a_1,1) - \ldots - (a_{n-1}, n-1) - (v,n)$ must be a path in G'.

Since v, $a_1, a_2, \ldots, a_{n-1}$ are mutually different, E(v,n) = E', E($a_i$, i) = E'－{e | e∈E' and e is associated with ($a_i$,1), …,( $a_i$, i-1)}, 1≤i≤n-1, therefore, we have E(v,n) contains $(v,0) - (a_1,1) - \ldots - (a_{n-1}, n-1) - (v,n)$, E($a_i$, i) contains $(v,0) - (a_1,1) - \ldots - (a_i, i)$, where, 1≤i≤n-1. This means $(v,0) - (a_1,1) - \ldots - (a_{n-1}, n-1) - (v,n)$ is a simple path in G'.

On the other side, if $(v,0) - (a_1,1) - \ldots - (a_{n-1}, n-1) - (v,n)$ is a simple path in G', $v - a_1 - a_2 - \ldots - a_{n-1} - v$ must be a path which is from v to v in G.

Since $(v,0) - (a_1,1) - \ldots - (a_{n-1}, n-1) - (v,n)$ is a simple path, we know that v, $a_1, a_2, \ldots, a_{n-1}$ are mutually different. This means that $v - a_1 - a_2 - \ldots - a_{n-1} - v$ is a Hamilton circle of G. ∎

**Theorem 5**  Let G be a undirected graph of order n. The complexity of transforming G into G' is a polynomial function of n.

**Proof:** For vertex u in G, step 3 will generate n-1 vertices in G'. Hence step 3 will generate (n-1)*(n-1) vertices. For all edges in G, step 4 will generate 2*n edges in G'. Hence step 4 will generate $2n^3$ edges at most. We can finish step 5 in $O(n^5)$, since E(u, $l$) have $2n^3$ edges at most and the number of E(u, $l$) is no more than $n^2$. The complexity of the algorithm is $O(n^5)$. ∎

## 5.  Conclusions

From all the discussion above, we have the following conclusion.

**Theorem 6** There exists a polynomial time algorithm to solve MSP problem. There exists a polynomial time algorithm to solve Hamilton circuit problem.

The most difficult thing in this paper is to prove that we can claim the existence of a simple path in G if Comp(E(D), D, R(E))≠∅ (that is Theorem 3). To do that, we proved Lemma 1, 2, 3, 4, 5 firstly. Then we proved **αβ lemma** based on Lemma 1, 2, 3, 4, 5. And finally we used **αβ lemma** to prove Theorem 3.

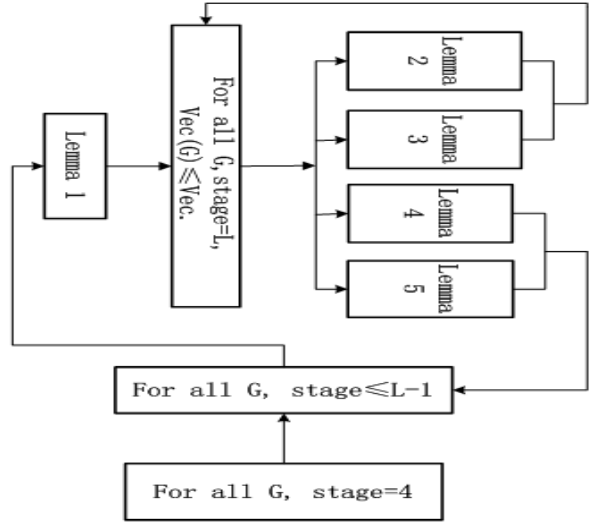

Figure 11 The Proving Logic for the Proving Algorithm

The logic relation among Lemma 1, 2, 3, 4, 5 and **αβ lemma** is shown in Fig. 11, which can be described as follows:

We directly verify that the Proving Algorithm makes correct assertion for all multistage graphs with four stages. Then we assume that the Proving Algorithm makes correct assertion for all multistage graphs with L-1 stages. For all graphs with L stages, if the Proving Algorithm is incorrect for some multistage graphs of this kind, we can find out the 'smallest' one with the linear order "≤" which we defined. Therefore, without losing generality, we can further assume that G is the 'smallest' graph that makes the Proving Algorithm fail in determining the existence of the said simple path in step 10. Then we will get a contradiction with the smallest graph, by splitting G to get a smaller graph (Lemma 2 and 3) or compressing G to get a graph with L-1 stages (Lemma 4 and 5).

To test Z-H algorithm, we need to generate instances of MSP. To generate an instance of MSP, we need to assign a set of edges to each E(v). E(v) is a subset of E, hence for all e∈E, we randomly decide whether e can be an edge in E(v) depending on the value of the current system time, i.e., e∈E(v) if and only if the value is odd.

We also need a creditable algorithm to tell us if the generated instance contains a simple path. Hence our testing system has three parts: the instance generator, the backtracking algorithm as a benchmark and Z-H algorithm. Until now, since 2010.10.06, more than 52 millions of instances have been generated randomly, each of which has 100 vertices. Some instances contain a simple path while others (it is the majority in all the generated instances) do not. All the results show that our polynomial time algorithm can get the same answer as the backtracking algorithm does. No exception.